\documentclass[fleqn,usenatbib]{mnras}

\usepackage[T1]{fontenc}
\usepackage{ae,aecompl}

\usepackage{graphicx}	
\usepackage{amsmath}	
\usepackage{amssymb}	

\title[Radio morphology and variability of Mrk\,590]{Parsec-scale radio morphology and variability of a changing-look AGN: the case of Mrk\,590}

\author[J. Y. Koay et al.]{
J. Y. Koay,$^{1}$\thanks{E-mail: koayjy@dark-cosmology.dk}
M. Vestergaard,$^{1,2}$
H. E. Bignall,$^{3}$ 
C. Reynolds,$^{3}$ 
and B. M. Peterson$^{4,5}$
\\
$^{1}$Dark Cosmology Centre, Niels Bohr Institute, University of Copenhagen, 2100 Copenhagen \O, Denmark\\
$^{2}$Steward Observatory, University of Arizona, Tucson, AZ 85721, USA\\
$^{3}$CSIRO Astronomy and Space Science, Kensington, WA 6152, Australia\\
$^{4}$Department of Astronomy, The Ohio State University, OH 43210, USA\\
$^{5}$Center for Cosmology and Astroparticle Physics, The Ohio State University, OH 43210, USA\\
}

\date{Accepted XXX. Received YYY; in original form ZZZ}

\pubyear{2015}

\begin{document}
\label{firstpage}
\pagerange{\pageref{firstpage}--\pageref{lastpage}}
\maketitle

\begin{abstract}
We investigate the origin of the parsec-scale radio emission from the changing-look active galactic nucleus (AGN) of Mrk\,590, and examine whether the radio power has faded concurrently with the dramatic decrease in accretion rates observed between the 1990s and the present. We detect a compact core at 1.6\,GHz and 8.4\,GHz using new Very Long Baseline Array observations, finding no significant extended, jet-like features down to $\sim$1\,pc scales. The flat spectral index ($\alpha_{1.6}^{8.4} = 0.03$) and high brightness temperature ($T_{\rm b} \sim 10^{8}\,\rm K$) indicate self-absorbed synchrotron emission from the AGN. The radio to X-ray luminosity ratio of ${\rm log}(L_{\rm R}/L_{\rm X}) \sim -5$, similar to that in coronally active stars, suggests emission from magnetized coronal winds, although unresolved radio jets are also consistent with the data. Comparing new Karl G. Jansky Very Large Array measurements with archival and published radio flux densities, we find $46\%$, 34\%, and (insignificantly) 13\% flux density decreases between the 1990s and the year 2015 at 1.4\,GHz, 5\,GHz and 8.4\,GHz respectively. This trend, possibly due to the expansion and fading of internal shocks within the radio-emitting outflow after a recent outburst, is consistent with the decline of the optical-UV and X-ray luminosities over the same period. Such correlated variability demonstrates the AGN accretion-outflow connection, confirming that the changing-look behaviour in Mrk\,590 originates from variable accretion rates rather than dust obscuration. The present radio and X-ray luminosity correlation, consistent with low/hard state accretion, suggests that the black hole may now be accreting in a radiatively inefficient mode.

\end{abstract}




\section{Introduction}\label{introduction}

In `optical changing-look' active galactic nuclei (AGNs), broad emission lines in the optical-UV spectra disappear or emerge within a span of a decade or less, in conjuction with a significant dimming or brightening of the optical-UV and X-ray continuum emission \citep[e.g.,][]{lamassaetal15}. The broad emission lines are produced by high-velocity gas in the vicinity of the accreting supermassive black hole (typically light-weeks away), ionized by the continuum photons from the accretion disk \citep[e.g.,][]{peterson97}. Such changing-look behavior, although very rare, has long been observed in nearby, lower-luminosity AGNs, namely Seyfert and low-ionization nuclear emission-line region (LINER) galaxies \citep[e.g.,][]{tohlineosterbrock76,penstonperez84,cohenetal86,storchi-bergmannetal93,aretxagaetal99,shappeeetal14}. Since the recent discovery of changing-look behavior in a more luminous quasar, SDSS J015957.64$+$003310.5 \citep{lamassaetal15}, about a dozen changing-look quasars have been uncovered at redshifts of $0.2 \lesssim z \lesssim 0.6$, either serendipitously \citep{runnoeetal16} or through systematic searches in multi-epoch Sloan Digital Sky Survey (SDSS) spectroscopic data \citep{macleodetal15,ruanetal15}.  

The variability of the optical-UV line and continuum fluxes may be attributed to variable extinction as clouds of gas and dust move across the line of sight, thereby obscuring or revealing the accretion disk and broad-line emitting region (BLR) in AGNs. Such a scenario is possible if the obscuring dusty torus posited to surround the BLR in standard AGN unification schemes \citep{urrypadovani95} has a clumpy structure \citep{nenkovaetal08, elitzur12}. Variable absorption by discrete clouds can explain spectral changes (between Compton-thin and Compton-thick states) observed on timescales of a decade in a small number of AGNs at X-ray wavelengths \citep[e.g.,][]{risalitietal05, marcheseetal12}, since the X-ray emission likely arises from a much more compact region \citep{risalitietal09} relative to the BLR. The estimated crossing-time of clouds in a Keplerian orbit around the more extended BLR can be dozens or hundreds of years in duration, and is therefore an unlikely cause of the variability on timescales of a decade or less observed in optical changing-look AGNs \citep{lamassaetal15,ruanetal15,runnoeetal16}. 

The dramatic changes in the spectral properties of changing-look AGNs are better modeled by intrinsic dimming (or brightening) due to changes in accretion properties, as opposed to dust extinction models \citep{lamassaetal15,macleodetal15,ruanetal15}. Disk instabilities can result in variable disk emission \citep{linshields86}, affecting the abundance of ionizing photons to excite the broad emission line gas. In disk-wind models of the BLR, changing accretion rates can alter the BLR structure, and at low bolometric luminosities may even lead to the disappearance of the BLR \citep{elitzurshlosman06,elitzuretal14}. In the case of SDSS J015957.64$+$003310.5, \citet{merlonietal15} argue that the decline in flux at X-ray and optical wavelengths is consistent with the power-law decay of a flare from a stellar tidal disruption event. Changing-look AGNs may even be transitioning between radiatively efficient and radiatively inefficient modes of accretion, analogous to spectral state transitions between high/soft accretion states and low/hard states observed in Galactic black hole binaries on timescales of months to years \citep[e.g.,][and references therein]{remillardmcclintock06}, and in the most extreme cases may even be transitioning between activity and quiescence. Therefore, changing-look AGNs provide an excellent laboratory for studying the accretion process in supermassive black holes using their variability within human lifetimes, previously thought possible only with Galactic black hole binaries.

These extreme variations in AGN accretion rates on such short timescales are surprising; AGNs are expected to turn on or off on timescales of hundreds of millions of years, as inferred by observations of a large population of low-luminosity AGNs with a wide range of levels of activity \citep[e.g., ][]{baldwinetal81, kewleyetal06}. By scaling the time-scale of spectral state transitions of Galactic black hole binaries to that of $\sim 10^8\,{\rm M_{\odot}}$ supermassive black holes, \citet{sobolewskaetal11} predict that state transitions in AGNs would have characteristic timescales of $10^5$ years. High resolution hydrodynamical simulations of AGN fueling in elliptical galaxies \citep{novaketal11} and gas-rich disc galaxies \citep{gaborbournaud13} reveal bursts of AGN activity with $10^5$ year and $10^6$ year timescales respectively, due to the clumpy structures in the inflowing gas. The AGNs are then expected to turn off on timescales of $10^4$ years, as inferred from observations of extended narrow emission-line regions in quiescent galaxies where the AGN has already faded \citep[e.g.,][]{schawinskietal10,keeletal15}. These timescales are orders of magnitude longer than that observed in changing-look AGNs, indicating that the mass inflow rates are not smooth on timescales of decades. Simple AGN unification models \citep[e.g.,][]{antonucci93,urrypadovani95}, in which the presence of broad emission lines in optical-UV spectra depends only on source-orientation, clearly need to be extended to include variable AGN luminosities and mass accretion rates.  

What is presently missing in the study of changing-look behaviour in AGNs, is the investigation of their possible radio variability. It is unknown if the strength of radio emission from changing-look AGNs, where present, vary in tandem with that of the optical-UV and X-ray emission. Investigating the relationships between the radio, X-ray, optical continuum, and emission line luminosities (and their variability) in changing-look AGNs can provide a better understanding of the disk-jet-wind-corona-BLR connection in accreting supermassive black holes. The observed tight correlation between radio and X-ray luminosities plus black hole masses in Galactic black hole binaries and AGNs form a fundamental plane of black hole activity \citep{merlonietal03, falckeetal04}. The physical origin of this empirical relationship is still not well-understood, due to the uncertain origin of the X-ray emission, as well as the radio emission in radio-quiet AGNs in which extended jet-like features are not observed. Possible sources of the unresolved radio emission in radio-quiet AGNs include low-power parsec-scale jets \citep[e.g.,][]{milleretal93,falckeetal96,nagaretal01}, disk winds \citep[e.g.,][]{blundellkuncic07,zakamskagreene14}, advection-dominated accretion flows \citep[ADAFs, ][]{mahadevan97,narayanetal98}, and the disk-coronae \citep{laorbehar08}. Understanding the physical coupling between the radio and X-ray emitting components in AGNs can thus also potentially shed light on their unification with stellar mass black holes, and the origin of radio emission in radio-quiet AGNs. 

\subsection{The changing-look Seyfert galaxy Mrk\,590}\label{clmrk590}

In this paper, we examine the radio properties and variability of Mrk\,590, a changing-look Seyfert galaxy at redshift $z = 0.0264$, to better understand the physical connections between the accretion disk, BLR and radio-emitting components in this AGN. The broad H$\beta$ emission line disappeared from its optical-UV spectra sometime between the years 2006 and 2012 \citep{denneyetal14}. Four decades of multi-wavelength (radio to X-ray) archival and published data enable the changing-look behavior in Mrk\,590 to be studied in more detail. Its black hole mass, a key parameter in the study of black hole accretion, is also reasonably well-determined from reverberation mapping to be $4.75 (\pm 0.74) \times 10^7\,{\rm M_{\odot}}$ \citep{petersonetal04}. Its proximity to us, compared to the more distant changing-look quasars, enables spatially resolved studies of its nuclear regions. While high angular resolution ALMA images do not detect any $^{12}$CO(3--2) line emission in the central 150\,pc, constraining the central $\rm H_2$ gas mass to $M_{\rm H_2}\lesssim 10^5\,M_{\odot}$ \citep{koayetal16}, there is still potentially enough gas to fuel the black hole for another $10^5\,\rm yr$ assuming Eddington-limited accretion (for $L_{\rm Edd} \approx 7 \times 10^{45} \, \rm erg\, s^{-1}$ and assuming a mass to light conversion efficiency of 0.1). Stronger constraints are needed to determine if the central engine is truly running out of gas. We have secured ALMA Cycle 3 observing time to obtain improved angular resolution observations of the higher CO transitions with the aim of mapping the gravitational potential, modeling the gas kinematics, and investigating the gas transport mechanisms (fueling and outflows) in this intriguing AGN. 

For reference, we provide a summary of the variability characteristics of Mrk\,590 here, as reported by \citet{denneyetal14} based on four decades of published optical-UV (both line and continuum) and X-ray fluxes:
\begin{enumerate}
\item The optical continuum and broad H$\beta$ emission line fluxes, measured to be $\approx 3 \times 10^{-16} \, \rm erg\, s^{-1}\,cm^{-2}\,$\AA$^{-1}$ and $\approx 5 \times 10^{-14} \, \rm erg\, s^{-1}\,cm^{-2}$ respectively in the 1970s, had increased by an order of magnitude by the early 1990s. 

\item At its highest observed bolometric luminosity in the 1990s ($L_{\rm bol} \approx 6 \times 10^{44} \, \rm erg\, s^{-1}$ as estimated from the rest-frame 5100\,\AA \,continuum flux), the black hole was accreting at $\sim 10$\% of the Eddington limit \citep{petersonetal04}. The optical continuum, UV continuum and H$\beta$ line fluxes, observed to be approximately $5 \times 10^{-15} \, \rm erg\, s^{-1}\,cm^{-2}\,$\AA$^{-1}$, $4 \times 10^{-14} \, \rm erg\, s^{-1}\,cm^{-2}\,$\AA$^{-1}$, and $5 \times 10^{-13} \, \rm erg\, s^{-1}\,cm^{-2}$ respectively in the 1990s, have since decreased by about two orders of magnitude in strength. In fact, the optical-UV continuum emission has faded to the point where it can now be modeled solely by emission from the host galaxy stellar populations. The total (both soft and hard) X-ray flux has also decreased by at least an order of magnitude, with the 0.5--10\,keV flux in 2013 measured at $1.3 \times 10^{-12} \, \rm erg\, s^{-1}\,cm^{-2}$. The narrow [O\,\textsc{iii}]\,$\lambda5007$ line flux has decreased by only a factor of two (from $1.1 \times 10^{-13} \, \rm erg\, s^{-1}\,cm^{-2}$ to $0.56 \times 10^{-13} \, \rm erg\, s^{-1}\,cm^{-2}$) within the same period, and is delayed by up to $\sim$10 years, reflecting the larger spatial extent of the emitting region.

\item It is unlikely that the observed changes in Mrk\,590 result from variable dust extinction. This is supported by the observed variability of the [O\,\textsc{iii}]\,$\lambda5007$ line and the fact that the observed variability timescales are shorter than the dynamical timescales required for an occulting cloud.

\end{enumerate}

It is unknown if the radio emission from Mrk\,590 also varied in strength during the initial brightening and eventual fading of the AGN. Published radio images of Mrk\,590 at frequencies of 1.4 to 8.4\,GHz, and at angular resolutions of tens of arcseconds ($\sim 5$ kpc) to 0.1 arcseconds ($\sim 50$ pc) reveal the presence of only a single unresolved (or marginally resolved) core component down to the image sensitivity limits \citep[e.g.,][]{theanetal01,schmittetal01}. The origin of this radio emission is unknown. Higher resolution images from very long-baseline interferometry (VLBI) are needed to determine the radio morphology of Mrk\,590 at pc-scales, allowing us to also probe its radio emission on spatial scales relevant to the timescale of the variability observed at other wavelengths. 

\subsection{Goals of this study}\label{thisstudy} 

In this study, we present the first investigation into the radio variability of a changing-look AGN over the period of its transition, using our recent multi-frequency observations of Mrk\,590 with the Karl G. Jansky Very Large Array (VLA, program ID: VLA/15A-084), along with available archival and published radio data of Mrk\,590 over the past four decades. We also present the highest angular resolution radio images of Mrk\,590, which we obtained using the Very Long Baseline Array (VLBA, program ID: BK192), to probe its parsec-scale radio morphology. The goals of this study are (1) to determine the physical origin of the radio emission in Mrk\,590 based on its spectral shape, brightness temperature and pc-scale morphology, (2) to establish if the radio emission from Mrk\,590 has varied in intensity in relation to that at optical-UV and X-ray wavelengths, and if so, to determine the possible causes of this variability, and quite importantly, (3) to obtain baseline data for the radio fluxes and pc-scale radio structures of Mrk\,590 for future comparisons, in the event that the AGN turns on again in the near future. In Section~\ref{obsdata}, we describe the new VLA and VLBA observations and data processing, as well as the ancillary archival and published data used for this study. In Section~\ref{results}, we present our results and discuss the origin of the radio emission in Mrk\,590, its variability, and their implications for changing-look AGNs. We adopt the following cosmology: $\Omega_m = 0.3$, $\Omega_\Lambda = 0.70$, and $H_0 = 70\,{\rm km\,s^{-1}\,Mpc^{-1}}$. This gives us a luminosity distance of $D_{\rm L} = 115.4\,{\rm Mpc}$ and a linear scale of 531\,pc per arcsec for Mrk\,590. 

\section{Observational Data}\label{obsdata}

Our preliminary inspection of published radio data of Mrk\,590 obtained in the 1980s and 1990s revealed that the radio flux densities may have varied over the past few decades. Motivated by these hints of radio variability, we obtained new VLA observations, described in Section~\ref{VLAobs}, to confirm our hypothesis that they follow trends consistent with the variations observed at higher energies. We discuss the observations and data processing for our VLBA data in Section~\ref{VLBAobs}, and present all the ancillary archival and published data used in our study in Section~\ref{ancdata}.

\subsection{New VLA observations and data processing}\label{VLAobs}

Our new observations of Mrk\,590 were carried out with 27 antennas on the VLA in A-configuration on 2015 June 23, using the P-band, L-band, C-band, X-band and Ku-band receivers (see Table~\ref{obsummary} for corresponding frequencies). We used the standard wideband continuum mode of the WIDAR correlator, selecting 8-bit sampling and 2 second correlator integration times for all receiver bands. In this mode, there are two basebands that can be tuned separately, thereby allowing each receiver to observe at two separate centre frequencies within the band. Observations of Mrk\,590 at each frequency band were bounded by (and at P-band and Ku-band also interleaved with) scans of the complex gain calibrator, J0215$-$0222. A single scan of 3C\,48 was also obtained at each frequency band for bandpass and flux calibration. For the higher-frequency Ku-band observations, we performed reference pointing scans on 3C\,48 to calibrate for pointing errors. The total on-source integration times, observing bandwidths, and centre frequencies for each baseband are presented in Table~\ref{obsummary}. 

\begin{table*}
\centering
\caption{Summary of the observational setup and the imaging parameters for the new 2015 VLA dataset\label{obsummary}}
\begin{tabular}{lccccc}
\hline
\hline

Band & Frequency$^a$ & Bandwidth  & On-source Time & Pixel size & Image size \\ 
		& (GHz)                & (GHz) & (min) & (arcsec) & (pixels)  \\
\hline
P & 0.38 & 0.256 & 16.9 & $1.0 \times 1.0 $ &$12000 \times 12000$\\
L & 1.3 & 0.512 & 4.1 & $0.2 \times 0.2$ &$12000 \times 12000$\\
  & 1.8 & 0.512 & 4.1  & $0.2 \times 0.2$ &$12000 \times 12000$\\
C & 5.0 & 1.024 & 4.1 & $0.06 \times 0.06$ & $9000 \times 9000$\\
   &  6.5 & 1.024 & 4.1 & $0.06 \times 0.06$ & $9000 \times 9000$\\
X & 8.6 & 1.024 & 5.0 & $0.04 \times 0.04$ & $7200 \times 7200$\\
   & 10.0 & 1.024 & 5.0 & $0.04 \times 0.04$ & $7200 \times 7200$\\
Ku & 15.0 & 1.024 & 8.5 & $0.025 \times 0.025$ & $7200 \times 7200$\\
    & 17.0 & 1.024 & 8.5 & $0.025 \times 0.025$ & $7200 \times 7200$\\

\hline
\end{tabular}
\begin{flushleft}
$^a$The centre frequencies for each of the two independently tuned basebands within each receiver band. At P-band, we combined both 128\,MHz basebands into a single, contiguous 256\,MHz band for better overall sensitivity.\\
\end{flushleft}
\end{table*}   

We performed the data processing using the Common Astronomy Software Applications (CASA) package \citep[version 4.4.0;][]{mcmullinetal07}. Our observations at 1.3\,GHz to 17\,GHz are straightforward, so we made use of the CASA VLA scripted calibration pipeline to perform standard bandpass, complex gain and flux calibration, including automated flagging of radio frequency interference (RFI). Hanning smoothing was applied to reduce the effects of spectral ringing resulting from narrowband RFI. We examined the pipeline-calibrated data and calibration tables to check for bad calibration solutions. We then performed further manual flagging of residual RFI in the data. Where such residual RFI were found in the calibrator sources, we reran the calibration pipeline after flagging the bad data. 

Since the 0.38\,GHz data obtained from observations conducted together with that of other bands as part of a single scheduling block cannot at present be converted directly into CASA's .ms format, we first used the Obit \citep{cotton08} and Astronomical Image Processing System \citep[AIPS, release 31DEC16;][]{greisen03} software packages to convert the 0.38\,GHz data into UVFITS format, in the process applying the online flags to the data. We then used the \texttt{importuvfits} task in CASA to convert the UVFITS files into .ms format for data reduction in CASA. We corrected for the erroneously labeled polarizations (from circular to linear polarizations), and found some antennas with swapped polarization components due to receiver cabling errors\footnote{these known errors are described here:\\ https://casaguides.nrao.edu/index.php/P-band\_basic\_data\_reduction}, which we also corrected for. RFI in the calibrator sources J0215$-$0222 and 3C\,48 were flagged manually, after which we performed the bandpass, complex gain and flux calibrations. We then inspected the data of Mrk\,590 for RFI, which we then flagged. 

Imaging and deconvolution were both performed using the CASA task \texttt{clean}. With the exception of the 0.38\,GHz data, imaging was carried out with natural weighting of the visibilities to obtain the best signal-to-noise ratio, as well as to obtain beam sizes comparable to that of published data at earlier epochs. For the 0.38\,GHz image, we used Briggs weighting (robustness = 0.5) to obtain a reasonable balance between surface brightness sensitivity and angular resolution ($\sim 5 ''$) for comparisons with the higher frequency data, since the largest angular scale that can be probed at 17\,GHz in the A configuration is $\sim 5''$. We derived a separate image from the visibilities of each of the two independently tuned basebands of each receiver, except at 0.38\,GHz where we have derived a single image from the combined visibilities at both basebands. This enabled us to improve the signal-to-noise ratio of the resulting 0.38\,GHz image, where the total observing bandwidth is narrower and significantly more data (about half the total number of channels) had to be excised due to strong RFI. Pixel sizes and image sizes used for all the images produced are presented in Table~\ref{obsummary}. For the deconvolution, we applied a mask at the location of the central source, and cleaned components within the mask down to three times the image rms intensities. We note that bright sources (10\,mJy levels at 1.3\,GHz and 1.8\,GHz, as well as 100\,mJy levels at 0.38\,GHz) are present within the primary beam of the lower frequency images, requiring these sources to be included in the mask such that the deconvolution of the dirty beam is also applied to these other sources within the field of view. We then performed phase self-calibration of the visibilities at each frequency band using the model visibilities of the clean components.   

For all data from 1.3\,GHz to 17.0\,GHz, we generated an alternate set of images applying a uv-tapering of 5$''$, so that the restoring beams for all images (including that at 0.38\,GHz) are of comparable sizes. This provides a way of checking for biases in flux measurements arising from decreasing image beam sizes with increasing observing frequency, when examining the spectral shape of Mrk\,590.

At all frequencies, we observe a core component in the centre of Mrk 590 (Figure 1), which we assume to be the location of the AGN. This component is unresolved (with visibilities consistent with that of a point source) at some frequencies, and is marginally resolved in others, as ascertained by fitting a source model with a Gaussian profile to the visibilities and/or images. We used the CASA task \texttt{imfit} to fit a two-dimensional Gaussian function to the central emission component of Mrk\,590 to extract the integrated flux densities, $S_{\rm int}$, and peak fluxes, $S_{\rm peak}$, at each observed frequency (presented in Table~\ref{VLAnewfluxes}). This was done for both the naturally weighted and uv-tapered images. 

\begin{figure*}
\begin{center}
\includegraphics[width=\textwidth]{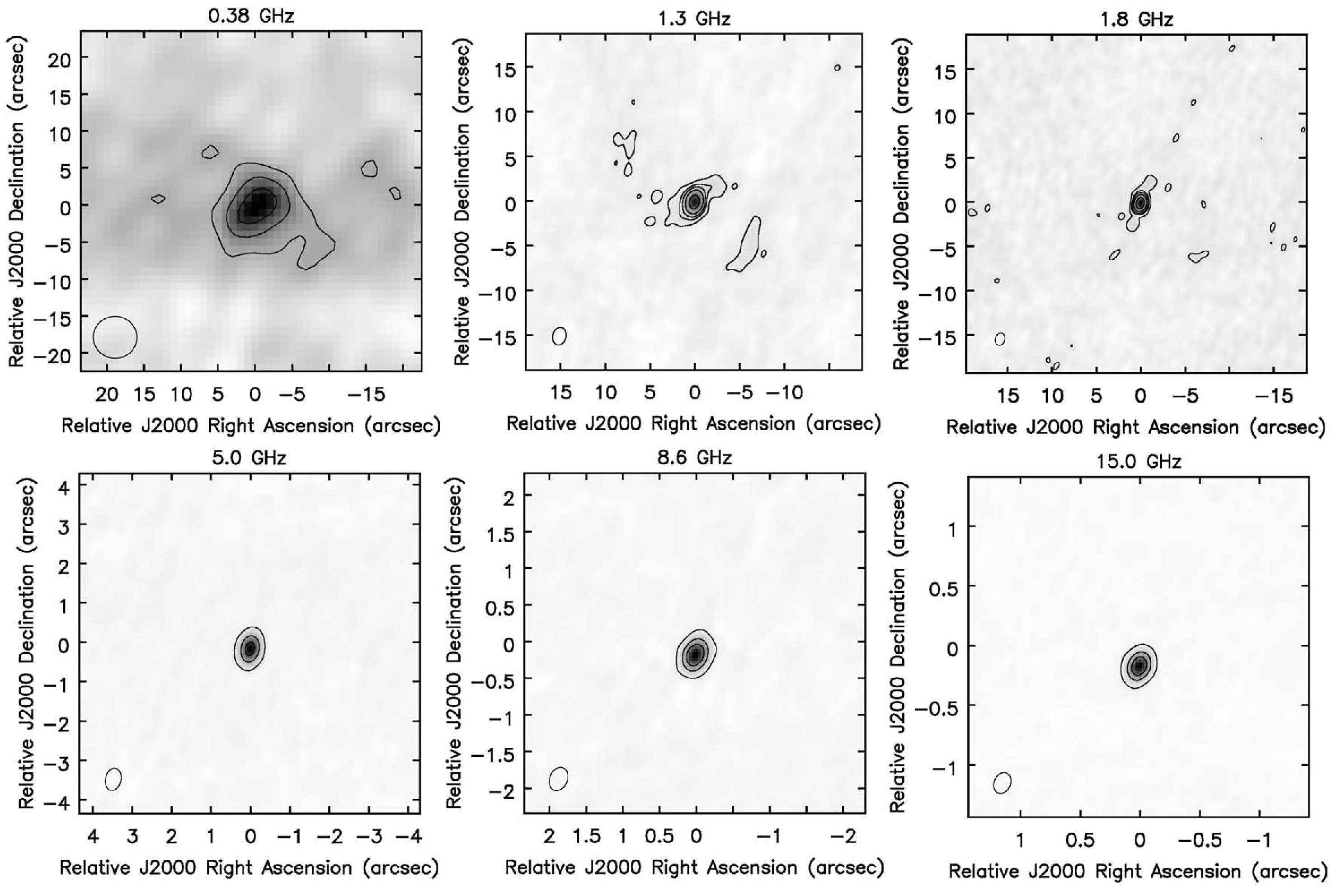}
\end{center}
\caption{{New VLA images of Mrk\,590 at 0.38\,GHz, 1.3\,GHz, 1.8\,GHz, 5.0\,GHz, 8.6\,GHz, and 15.0\,GHz shown as greyscale maps and black contours. The image rms intensities are 0.54\,mJy\,beam$^{-1}$, 0.061\,mJy\,beam$^{-1}$, 0.042\,mJy\,beam$^{-1}$, 0.021\,mJy\,beam$^{-1}$, 0.016\,mJy\,beam$^{-1}$, and 0.015\,mJy\,beam$^{-1}$, respectively. Contour levels are: 5$\sigma$, 10$\sigma$, and 15$\sigma$ at 0.38\,GHz; 4$\sigma$, 8$\sigma$, 12$\sigma$, 20$\sigma$, and 40$\sigma$ at 1.3\,GHz and 1.8\,GHz; 10$\sigma$, 50$\sigma$, and 100$\sigma$ at 5.0\,GHz, 8.6\,GHz and 15.0\,GHz. All images are obtained using natural weighting, except at 0.38\,GHz where Briggs weighting (robustness $= 0.5$) is used. The FWHM clean beam sizes (shown at the bottom left of each image) are presented in column~2 of Table~\ref{VLAnewfluxes}. Coordinates are relative to the phase centre at R.A.: $02^{\rm h}14^{\rm m}33 \fs 560$ and DEC: $-00^{\circ}46'00\farcs 00$. Images at centre frequencies of 6.5\,GHz, 10.0\,GHz and 17.0\,GHz contain only an unresolved or marginally resolved core component (similar to the 5.0\,GHz, 8.6\,GHz and 15.0\,GHz images), and are thus not shown.} \label{vlaimages}}
\end{figure*}

\begin{table*}
\centering
\caption{Multi-frequency image properties and source fluxes for Mrk\,590 from new VLA observations in 2015. \label{VLAnewfluxes}}
\begin{tabular}{lccccccc}
\hline
\hline

Frequency & Beam size$^a$ & Beam P.A. & ${S_{\rm int}}^b$ & ${S_{\rm peak}}^b$ & ${S_{\rm int}(5'')}^b$ & ${S_{\rm peak}(5'')}^b$ \\ 
(GHz) & ($'' \times ''$) & ($^{\circ}$)  & (mJy) & (mJy beam$^{-1}$) & (mJy) & (mJy beam$^{-1}$)\\
\hline
0.38 & $5.90\times 5.66$ & $+$86.7  & 16.00 $\pm$ 1.400(1.61) & 9.080 $\pm$ 0.540(0.71) & -- & -- \\
1.3 & $1.99 \times 1.46$ & $-$12.8 &  3.390 $\pm$ 0.110(0.20) & 3.246 $\pm$ 0.061(0.17) & 4.76 $\pm$ 0.20(0.31) & 4.35 $\pm$ 0.11(0.24)\\
1.8  & $1.43 \times 1.06$ & $-$15.9 &  2.754 $\pm$ 0.075(0.16) & 2.724 $\pm$ 0.042(0.14) & 3.56 $\pm$ 0.20(0.27) & 3.29 $\pm$ 0.11(0.20)\\ 
5.0  & $0.57 \times 0.39$ & $-$11.9 &  2.770 $\pm$ 0.037(0.14) & 2.723 $\pm$ 0.021(0.14) & 3.58 $\pm$ 0.11(0.21) & 3.136 $\pm$ 0.059(0.17)\\ 
6.5  & $0.42 \times 0.29$ & $-$16.9 &  2.902 $\pm$ 0.035(0.15) & 2.815 $\pm$ 0.019(0.14) & 3.56 $\pm$ 0.16(0.24) & 3.205 $\pm$ 0.087(0.18)\\ 
8.6  & $0.33 \times 0.24$ & $-$25.7 &  3.112 $\pm$ 0.029(0.16) & 3.023 $\pm$ 0.016(0.15) & 3.84 $\pm$ 0.14(0.24) & 3.479 $\pm$ 0.078(0.19)\\ 
10.0 & $0.28 \times 0.21$ & $-$27.6 & 3.064 $\pm$ 0.029(0.16) & 2.964 $\pm$ 0.016(0.15) & 3.40 $\pm$ 0.10(0.20) & 3.307 $\pm$ 0.055(0.17)\\ 
15.0  & $0.18 \times 0.14$ & $-$23.6 & 2.666 $\pm$ 0.027(0.14) & 2.583 $\pm$ 0.015(0.13) & 2.65 $\pm$ 0.07(0.15) & 2.643 $\pm$ 0.039(0.14)\\ 
17.0  & $0.16 \times 0.12$ & $-$22.1 & 2.470 $\pm$ 0.033(0.13) & 2.418 $\pm$ 0.018(0.12) & 2.55 $\pm$ 0.11(0.17) & 2.614 $\pm$ 0.065(0.15)\\ 
\hline
\end{tabular}
\begin{flushleft}
$^a$The full width at half maximum (FWHM) of the synthesized clean beam for the naturally weighted images, except at 0.38\,GHz where Briggs weighting (robustness = 0.5) is used.\\
$^b$The uncertanties outside the parentheses are the stochastic uncertainties, $\sigma_{\rm stat}$, which in the case of $S_{\rm peak}$ is the image rms intensity. The total estimated uncertainties inclusive of systematic calibration errors, $\sigma_{\rm tot}$, are quoted within the parentheses. 
\end{flushleft}
\end{table*} 

The estimated total uncertainties in the values of $S_{\rm int}$ and $S_{\rm peak}$, including systematic calibration errors, are shown in parentheses in Table~\ref{VLAnewfluxes}. As the probability distribution of the total error, $\sigma_{\rm tot}$, is the convolution of the probability distribution of the flux independent stochastic errors and that of the flux dependent calibration errors, the total uncertainties are estimated as:
\begin{equation}\label{sigmatot} 
\sigma_{\rm tot} = \sqrt{\sigma_{\rm stat}^2 + \sigma_{\rm cal}^2} 
\end{equation}
where $\sigma_{\rm stat}$ represents the uncertainties in the fitting of the Gaussian function to the image (and thus includes the image rms noise). $\sigma_{\rm cal}$ represents the systematic calibration errors, dominated by the uncertainties in the absolute flux density scaling of our flux calibrator. We assume very conservatively that the calibration errors have values that are 5\% of the measured fluxes, such that $\sigma_{\rm cal} = 0.05S_{\rm int}$ for the integrated flux densities and $\sigma_{\rm cal} = 0.05S_{\rm peak}$ for the peak fluxes. We also note that our flux calibrator for our new VLA observations, 3C\,48, has been observed to exhibit flux variations of up to $\sim$5 percent at frequencies of 1.4\, GHz to 8.4\,GHz, and up to $\sim 10$ percent at 15\,GHz, over the last 30 years \citep{perleybutler13}. However, since we use the \citet{perleybutler13} scheme for flux density scaling, which has accurate flux measurements of 3C\,48 up to the year 2012, our flux scaling errors are unlikely to be this large. This is because the flux variations in 3C\,48 are typically $< 3\%$ between 1.4\,GHz and 8.4\,GHz (and $< 5\%$ at 15\,GHz) on two to three year timescales, so the flux levels at these frequencies are unlikely to have changed by more than that amount between the years 2012 and 2015. 

\subsection{New VLBA observations and data processing}\label{VLBAobs}
	
VLBA observations at two different frequencies, centred at 1.6\,GHz and 8.4\,GHz, were carried out in 2015 February 3 with nine antennas (Hancock station did not participate in the observations due to snowstorms). At each frequency, we used dual-polarization observations, with a bandwidth of 256\,MHz per polarization and a 2-bit sampling rate, acquiring an aggregate recording bit rate of 2048\,Mbps. We used a correlator averaging time of 2 seconds. Observations of Mrk\,590 were carried out in phase referencing mode, where the quasar J0219$+$120 (located $2.4^{\circ}$ away from Mrk\,590) was used as our phase reference calibrator. We also observed 3C\,84 as our fringe finder and bandpass calibrator. We achieved a total observing time of 36 minutes on-source for Mrk\,590 at each frequency.

Processing of the VLBA data was carried out using the procedures defined in \texttt{VLBAUTIL} in the AIPS software package (release 31DEC16). This release of AIPS includes modified and new procedures that enable more accurate amplitude calibrations. It reduces the 20\% to 30\% discrepancies in the flux scaling (the cause of which at present is still not fully understood) found in recent VLBA observations that use the Polyphase Filter-bank (PFB) personality of the Roach Digital Backend (RDBE) system, as described by \citet{walker14}. We first inspected the visibilities to search for spurious data and RFI, which we flagged. We then performed the preliminary calibrations, correcting for Earth orientation parameters, ionospheric dispersive delays, digital sampling effects, instrumental delays and the bandpass responses. Amplitude calibration was performed using the procedure \texttt{VLBAAMP}, based on the antenna gain tables and antenna system temperatures measured during the observations. Parallactic angle corrections were applied using \texttt{VLBAPANG}. Delay, rate and phase calibrations were carried out using the procedures \texttt{VLBAPCOR} and \texttt{FRING}. The calibrated data were exported into UVFITS format using the AIPS task \texttt{FITTP}, and were converted to .ms format using the \texttt{importuvfits} task in CASA for further imaging and processing in CASA.  

Imaging and deconvolution were performed using the CASA task \texttt{clean}. After exploring different weighting schemes, we implemented natural weighting for both the 1.6\,GHz and 8.4\,GHz images, for better surface brightness sensitivity. We used pixel sizes of 1 milliarcsecond $\times$ 1 milliarcsecond (mas) and an image size of 4096 pixels $\times$ 4096 pixels at 1.6\,GHz. For the image at 8.4\,GHz, we used a pixel size of 0.17 mas $\times$ 0.17 mas and an image size of 6400 $\times$ 6400 pixels. The clean model was used as an input model for phase self-calibration. We achieved an rms noise level of 0.08\,mJy\,beam$^{-1}$ at 1.6\,GHz and 0.09\,mJy\,beam$^{-1}$ at 8.4\,GHz. The clean beam sizes are 14\,mas $\times$ 6.3\,mas ($\rm P.A. = 9.5^{\circ}$) at 1.6\,GHz, and 2.8\,mas $\times$ 1.2\,mas ($\rm P.A. = -1.0^{\circ}$) at 8.4\,GHz. 

\begin{figure}
\begin{center}
\includegraphics[width=\columnwidth]{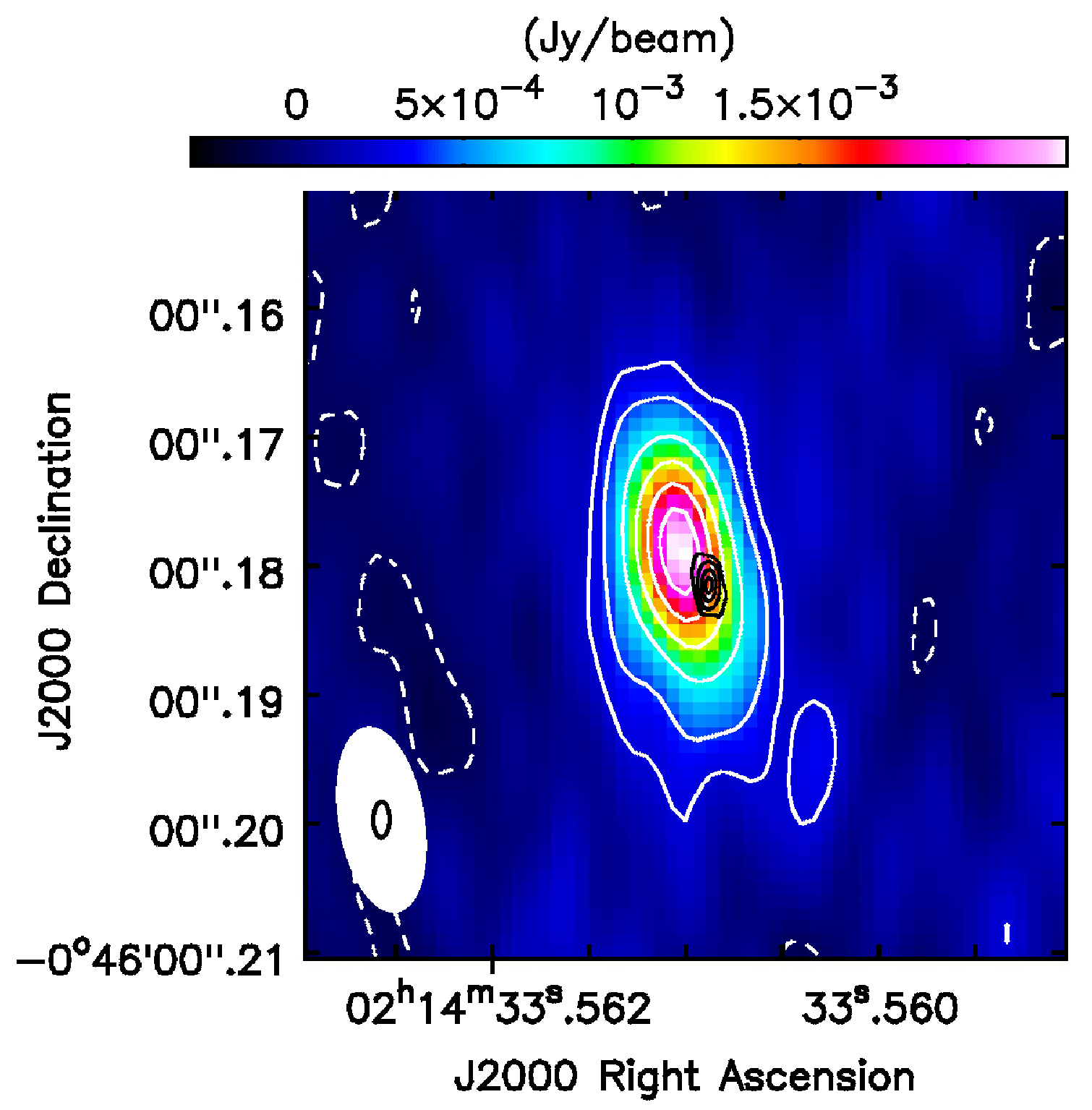}
\end{center}
\caption{{The VLBA images at 1.6\,GHz (colour map and white contours) and 8.4\,GHz (black contours), both obtained using natural weighting. White contours represent $-1\sigma$ (dashed), 3$\sigma$, 5$\sigma$, 10$\sigma$, 15$\sigma$, 20$\sigma$ and 25$\sigma$ levels in the 1.6\,GHz image, for a 1$\sigma$ rms intensity of $\rm 80\, \mu Jy\,beam^{-1}$. Black contours are at 5$\sigma$, 10$\sigma$, 20$\sigma$ and 30$\sigma$ levels (1$\sigma = 90\,\rm \mu Jy\,beam^{-1}$) in the 8.4\,GHz image. The FWHM synthesized beams of $\rm 14\,mas \times 6.3\,mas$ at 1.6\,GHz and $\rm 2.8\,mas \times 1.2\,mas$ at 8.4\,GHz are shown as the white filled ellipse and black ellipse, respectively, at the bottom left of the image.} \label{vlbaimage}}
\end{figure}

We detect a compact core component at 1.6\,GHz and 8.4\,GHz, and at 25$\sigma$ and 30$\sigma$ significance respectively (Figure~\ref{vlbaimage}). In the VLBA images at both frequencies, the sources are offset from the phase centre of the observations (R.A.: $02^{\rm h}14^{\rm m}33 \fs 559000$ and DEC: $-00^{\circ}46'00\farcs 27000$) by about 0.095 arcsec. This offset is sufficiently small such that the source is well within the field of view at both frequencies, limited by bandwidth smearing to be 11 arcsec at 1.6\,GHz and by time smearing to be 3.7 arcsec at 8.4\,GHz. For each of the core components at both frequencies, we present in Table~\ref{VLBAfluxes} their $S_{\rm int}$ and $S_{\rm peak}$ values, as well as their coordinates and estimated source sizes. These parameters were derived by fitting an elliptical Gaussian function to the images using the task \texttt{imfit} in CASA. Since we used the new VLBA calibration strategy as recommended by \citet{walker14}, which is expected to reduce the 20\% to 30\% VLBA flux discrepancy down to an error of 4\%, we again adopt a value of 5\% for $\sigma_{\rm cal}$, and include this term in our estimate of the overall $\sigma_{\rm tot}$ for our VLBA fluxes following Equation~\ref{sigmatot}.

\begin{table*}
\centering
\caption{Properties of the radio core detected in the 1.6\,GHz and 8.4\,GHz VLBA images of Mrk\,590. \label{VLBAfluxes}}
\begin{tabular}{lcccccccc}
\hline
\hline

Frequency & Weighting & ${S_{\rm int}}^a$ & ${S_{\rm peak}}^a$ & R.A. & DEC & ${\theta_{\rm maj}}^b$ & ${\theta_{\rm min}}^b$ & P.A.$^b$\\ 
(GHz) &   & (mJy) & (mJy beam$^{-1}$) & (J2000) & (J2000) & (mas) & (mas) & ($^{\circ}$)\\
\hline
1.6 & natural &  2.91 $\pm$ 0.18(0.23) & 2.06 $\pm$ 0.08(0.13) & $02^{\rm h} 14^{\rm m} 33 \fs 561020$ & $-00^{\circ} 46'00 \farcs 1795$ & 9.1 & 4.1 & 10.2\\
8.4  & natural &  3.04 $\pm$ 0.17(0.23) & 2.85 $\pm$ 0.09(0.17) &  $02^{\rm h} 14 ^{\rm m} 33 \fs 560879$ & $-00^{\circ} 46'00 \farcs 1815$ & $< 0.98$ & $< 0.28$ & --\\
\hline
\end{tabular}
\begin{flushleft}
$^a$The uncertanties outside the parentheses are the stochastic uncertainties ($\sigma_{\rm stat}$), while the total estimated uncertainties inclusive of systematic calibration errors, $\sigma_{\rm tot}$, are quoted within the parentheses.\\
$^b$ ${\theta_{\rm maj}}$, ${\theta_{\rm min}}$, and P.A. are the major axis, minor axis and parallactic angle of the core component as modeled by fitting an elliptical Gaussian function to the image.
\end{flushleft}
\end{table*} 

\subsection{Ancillary archival and published data}\label{ancdata}

To examine the radio variability of Mrk\,590 over the past four decades, we searched for and found published flux measurements of Mrk\,590 obtained from instruments such as the Westerbork Synthesis Radio Telescope (WSRT), the Multi-Element Radio Linked Interferometer Network (MERLIN), and the VLA in different array configurations from the 1970s to the present (shown in Table~\ref{archivaldata}). These include flux measurements of Mrk\,590 obtained from catalogues of radio surveys such as the NRAO VLA Sky Survey (NVSS) \citep{condonetal98}, the Faint Images of the Radio Sky at Twenty-cm (FIRST) Survey \citep{beckeretal95}, and the high-resolution VLA survey of the SDSS Stripe 82 (VLAS82) \citep{hodgeetal11}. 

\begin{table*}
\centering
\caption{Archival and Published Radio Continuum Data for Mrk\,590 
\label{archivaldata}}
\begin{tabular}{l c c c c c c c}
\hline
\hline
Frequency & Year & Instrument & Beam Size & ${S_{\rm int}}^a$ & ${S_{\rm peak}}^{a}$ & Recalibration$^b$ & Ref.$^{c}$ \\
  (GHz) & &   & ($'' \times ''$) & (mJy)  & (mJy\,beam$^{-1}$) & \\
\hline
1.4 & 1977 	&	WSRT & $^d$ & 11.0 $\pm$ 2.0(2.3) & -- & -- & 1 \\ 
	& 1983 & VLA-D & 90 $\times$ 90$^e$ & 11.2 $\pm$ 1.4(1.8) & -- & -- & 2 \\
	& 1983 & VLA-A & 1.95 $\times$ 1.35 & 4.86 $\pm$ 0.59(0.8) & 4.59 $\pm$ 0.33(0.57) & Y & 0\\
	& 1993 & VLA-D & 45 $\times$ 45$^e$ & 16.2 $\pm$ 0.6(1.7) & -- & -- &3\\
	& 1995 & MERLIN & 0.33 $\times$ 0.25 & 6.28 $\pm$ 0.22(0.67) & 6.05 $\pm$ 0.12(0.62) & Y & 4\\ 
	& 2002$^f$ & VLA-B   & 6.4 $\times$ 5.4 & 9.90 $\pm$ 0.50(1.11) & 6.83 $\pm$ 0.14(0.70) & -- & 5\\
	& 2008 & VLA-A  & 2.35 $\times$ 2.05 & 4.35 $\pm$ 0.10(0.45) & 2.92 $\pm$ 0.06(0.30) & -- & 6\\
\hline
4.9 & 1982 & VLA-A & -- & 3.5 $\pm$ 0.5(0.6) & -- & -- & 7 \\
      & 1983 & VLA-D & 15 $\times$ 15 & 5.23 $\pm$ 0.29(0.60) &  -- & -- & 2\\
      & 1995 & MERLIN & 0.33 $\times$ 0.25$^e$ & 4.23 $\pm$ 0.31(0.52) & 3.66 $\pm$ 0.16(0.40) & Y & 0\\ 
\hline 
8.4 & 1991 & VLA-A & 0.38 $\times$ 0.28 & 3.67 $\pm$ 0.14(0.39) & 3.32 $\pm$ 0.07(0.34) & Y & 8 \\
		& 1992 & VLA-C & 8.36 $\times$ 2.31 & 4.30 $\pm$ 0.32(0.54) &  3.39 $\pm$ 0.14(0.37) & Y &8 \\
		&1998 & VLA-A & 0.33 $\times$ 0.23  & 3.56 $\pm$ 0.06(0.37) & 3.38 $\pm$ 0.03(0.34) & Y & 9\\
\hline 
20 & 1983 & OVRO & 90 $\times$ 90 & 3.6 $\pm$ 0.6(0.7) &  -- & -- & 2\\
\hline
\end{tabular}
\begin{flushleft}
$^a$The uncertanties outside the parentheses are the stochastic uncertainties ($\sigma_{\rm stat}$), while the total estimated uncertainties inclusive of systematic calibration errors, $\sigma_{\rm tot}$, are quoted within the parentheses.\\
$^b$Y: indicates that the values of ${S_{\rm int}}$ and ${S_{\rm peak}}$ in that particular row are derived from data that we have recalibrated.\\
$^c$for data that we have recalibrated, the references show the papers in which the data was originally published\\
$^d$the source size was constrained to be $< 13''$ in the E-W direction\\
$^e$with uv-tapering applied\\
$^f$flux measurements derived from multiple images obtained over a period of 12 years, with 2002 being the mean year\\
References. -- (0)~unpublished archival data. (1)~\citet{wilsonmeurs82}. (2)~\citet{edelson87}. (3)~The NRAO VLA Sky Survey (NVSS) catalogue, \citet{condonetal98}. (4)~\citet{theanetal01}. (5)~The Faint Images of the Radio Sky at Twenty-cm (FIRST) Survey catalogue,  \citet{beckeretal95} (6)~\citet{hodgeetal11}. (7)~\citet{ulvestadwilson84}. (8)~\citet{kukulaetal95}. (9) \citet{kinneyetal00,schmittetal01}
\end{flushleft}
\end{table*}

We also searched for and obtained raw data from the VLA and MERLIN archives for reprocessing and recalibration, to serve as consistency checks. Using the CASA software package, we recalibrated the archival data observed with the VLA in A configuration (where available), since these are the most relevant for comparisons with that of our new VLA observations in the same configuration. Complex gain and flux calibrations, including phase self-calibration, were perfomed using standard methods. We found observations of Mrk\,590 at both 1.6\,GHz and 5\,GHz in the MERLIN archives, the latter of which (to our knowledge) appears to be unpublished. We recalibrated the MERLIN data using AIPS, applying standard methods. For all the recalibrated archival data, we derived the images using natural weighting. For the 5\,GHz MERLIN data, uv-tapering was applied  when obtaining the image such that the beam is comparable to the 1.6\,GHz MERLIN and 8.4\,GHz VLA A-configuration data obtained at a similar epoch (in the 1990s). The rows flagged with a `Y' in Table~\ref{archivaldata} indicate flux densities obtained from images derived from our recalibrated data. As with our new VLA and VLBA flux densities, we estimate the total uncertanties of $S_{\rm int}$ and $S_{\rm peak}$ using Equation~\ref{sigmatot}. Since the archival and catalogue data are obtained with different instruments and the flux densities are calibrated with different flux calibrators, we adopt, for each of these flux measurements, a more conservative calibration error of 10\% of the measured flux. 

For four entries in Table~\ref{archivaldata} for which we recalibrated the raw data and the flux densities are available in the literature, we find the published flux densities to be consistent with that derived from our recalibrated images to within 1$\sigma$ of both the stochastic and total uncertainties, with one exception. For the 8.4\,GHz VLA data from 1998, \citet{schmittetal01} obtain a flux density of 3.40$\pm$0.03\,mJy, at least 0.13\,mJy lower in flux (by twice the stochastic uncertainty of $\pm$0.06\,mJy) than our estimated value of 3.56\,mJy. However, the previously published value is well within our estimated total 1$\sigma$ uncertainty of $\pm$0.37\,mJy.

\section{Analyses and Discussion}\label{results}

\subsection{Kiloparsec to parsec scale radio morphology}\label{pcmorph}

Our new multi-frequency VLA images (Figure~\ref{vlaimages}) reveal a single, unresolved (or marginally resolved) compact radio core component at all observed frequencies. These are consistent with earlier published radio images of Mrk\,590 on these scales (references can be found in Table~\ref{archivaldata}). At both 1.3\,GHz and 1.8\,GHz, we also detect extended components at a radius of about 2\,arcsec ($\sim 1\,{\rm kpc}$) and 6\,arcsec ($\sim 3\,{\rm kpc}$) from the core. These extended components are spatially coincident with two ring-like structures of molecular gas at similar radii seen in $^{12}$CO(3--2) line images \citep{koayetal16}, likely associated with star formation processes within the ring. We do not discuss these extended components further in the present paper. 

In both of the VLBA images (Figure~\ref{vlbaimage}), we detect a single core component; fitting a Gaussian profile to the images reveals that the core emission is marginally resolved at 1.6\,GHz and unresolved at 8.4\,GHz (beam deconvolved source sizes, or upper limits thereof, are shown in Table~\ref{VLBAfluxes}). We confirm that the sub-parsec and parsec scale core components, detected by the VLBA at 8.4\,GHz and 1.6\,GHz respectively, fully account for the flux densities of the unresolved core radio emission as measured by the lower angular resolution VLA observations at these frequencies. Our new VLA and archival A-configuration VLA core flux densities above 1.6\,GHz are thus likely dominated by emission from the sub-pc to pc scale core components. On 1 mas to 10 mas scales (corresponding to linear scales of $\sim 0.5$\,pc and 5\,pc), we still do not detect the presence of extended jet-like structures in Mrk\,590.

\subsection{Origin of radio emission in Mrk\,590}\label{radioorigin}

We now examine the origin of the radio core emission in Mrk\,590 based on the radio spectral indices, brightness temperatures and ratios of the radio to X-ray luminosities. 

\citet{theanetal01} estimate the spectral index of Mrk\,590 to be $\alpha_{1.6}^{8.4} = -0.32$ (for $S_{\nu} \propto \nu^{\alpha}$) between 1.6 and 8.4\,GHz, using flux measurements separated by three years in the 1990s. This is consistent with $\alpha_{1.4}^{4.9} = -0.32$ estimated by \citet{edelson87} using concurrent measurements in 1983 but with a much larger 90 arcsec beam. 

\begin{figure}
\begin{center}
\includegraphics[width=\columnwidth]{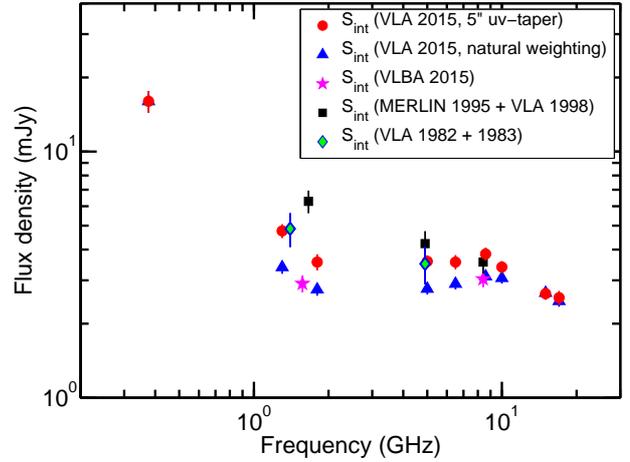}
\end{center}
\caption{The radio spectral energy distribution of Mrk\,590 at different epochs, using only flux densities from VLA observations in A-configuration (both archival and new), plus MERLIN and VLBA observations. We note that the VLA flux densities were obtained from images with frequency dependent beam-sizes. However, the VLBA fluxes (magenta stars) are comparable to that measured with the VLA in A-configuration in 2015 (blue triangles), indicating that the VLA fluxes are dominated by the emission from the central few parsecs, such that the effects of the different beam sizes are negligible above 1.4\,GHz. We observe a decrease in the flux densities between the 1990s (black squares) and the year 2015 (blue triangles), most notably at 1.4\,GHz. {\label{radiosed}}}
\end{figure}

From our new near-simultaneous VLA observations with better frequency sampling, we obtain a steep spectral index of $\alpha_{0.3}^{1.3} = -0.8$ between 0.3\,GHz and 1.3\,GHz (Figure~\ref{radiosed}). This steep spectrum is consistent with non-thermal, optically thin synchrotron emission, most likely originating from supernova remnants in the host galaxy, where the beam at 0.3\,GHz encompasses the inner star-forming molecular ring of $\approx$2 arcsec radius detected by \citet{koayetal16}. 

Between 1.8\,GHz and 8.4\,GHz, the spectral index flattens to $\alpha_{1.8}^{8.6} = 0.02$ (Figure~\ref{radiosed}), based on the 2015 VLA integrated flux densities. From the VLBA flux densities, we obtain $\alpha_{1.6}^{8.4} = 0.03$. We also estimate the rest-frame brightness temperatures from the VLBA flux densities, using the following equation for elliptical sources \citep{ulvestadetal05}:
\begin{equation}\label{btemp} 
T_{\rm b} = 1.8 \times 10^9 (1+z) \left( \dfrac{S_{\nu}}{\rm 1\,mJy}\right) {\left( \dfrac{\nu}{\rm 1\,GHz}\right)}^{-2} {\left(\dfrac{\theta_{\rm maj} \theta_{\rm min}}{\rm mas^2}\right)}^{-1}\,\rm K
\end{equation}
where $\theta_{\rm maj}$ and $\theta_{\rm min}$ are the major and minor axes of the source, as estimated from the fit of the elliptical Gaussian function to the core component. From the source sizes in Table~\ref{VLBAfluxes}, we obtain $T_{\rm b} = 6 \times 10^7\,{\rm K}$ at 1.6\,GHz and $T_{\rm b} = 3 \times 10^8\,{\rm K}$ at 8.4\,GHz. The brightness temperatures of HII regions and supernova remnants in star forming regions do not exceed $\rm 10^{5}\,K$ \citep{condon92}, so the emission must originate from the AGN itself. Additionally, if one assumes that the total 1.6\,GHz VLBA flux density is dominated by supernova remnants, it implies a star formation rate of ${\rm SFR}(M_* \geq 5\, M_{\odot}) \approx 1\,M_{\odot}\,\rm yr^{-1}$ (following \citet{condon92}). This inferred star formation rate far exceeds the upper limit of $\sim 10^{-4}\,M_{\odot}\,\rm yr^{-1}$ estimated based on constraints of the $\rm H_{2}$ gas mass in the central 150\,pc of Mrk\,590 \citep{koayetal16}.

In its present accretion state, the low bolometric luminosity of Mrk\,590 ($L_{\rm bol}/L_{\rm Edd} \sim 10^{-3}$, Table~\ref{Xrayfluxes}) suggests that the black hole may be accreting via advection dominated accretion flows \citep[ADAFs,][]{narayanetal98}, which are radiatively inefficient. However, thermal emission from ADAFs can be ruled out as the dominant source of the radio emission, due to the high brightness temperatures, as discussed above. Additionally, the radio luminosity of thermal emission from ADAFs can be estimated as \citep{yiboughn98}:
\begin{equation}\label{radiolumadaf} 
L_{\rm R} = 7 \times 10^{35} \left( \dfrac{M_{\rm bh}}{10^{7}\,M_{\odot}}\right) \left( \dfrac{L_{\rm X}}{\rm 10^{40}\,erg\,s^{-1}}\right)^{0.1} \left( \dfrac{\nu}{15\,{\rm GHz}}\right)\,\,{\rm erg\,s^{-1}}
\end{equation}
which predicts a 8.4\,GHz radio luminosity of $L_{\rm R} \sim 10^{36} \,{\rm erg\,s^{-1}}$ for an observed X-ray luminosity (2--10\,keV) of $L_{\rm X} = 2.6 \times 10^{42} \,{\rm erg\,s^{-1}}$ (Table~\ref{Xrayfluxes}). This is two orders of magnitude lower than our observed VLBA 8.4\,GHz radio core luminosity of $\sim 10^{38} \,{\rm erg\,s^{-1}}$. 

Free-free emission from an extended accretion disk or the ionized inner edge of the dusty torus, as seen in VLBI images of the obscured AGN NGC\,1068 \citep{gallimoreetal97}, is also ruled out based on the high $T_{\rm b}$ values in Mrk\,590.

The high brightness temperature of $10^8\,{\rm K}$ at 8.4\,GHz suggests a non-thermal origin of the radio core emission. The flat spectrum between 1.8\,GHz and 8.4\,GHz is consistent with self-absorbed synchrotron emission from multiple optically thick components. Although the estimated brightness temperatures are not as high as required ($T_{\rm b}\rm \gtrsim 10^9\,K$) for synchrotron self-absorption, we note that our estimated $T_{\rm b}$ values are lower bounds due to the limited VLBA angular resolution. To be consistent with synchrotron self-absorption, the total $\approx 3\,{\rm mJy}$ flux density at 8.4\,GHz must originate from a region no larger than 0.1\,pc ($\sim 10^4$ gravitational radii) in size, such that $T_{\rm b}\rm \sim 10^9\,K$. This is consistent with radio source sizes of 0.05\,pc to 0.2\,pc inferred from the brightness temperatures of flat-spectrum radio cores in other nearby Seyfert galaxies \citep{mundelletal00}. This synchrotron-emitting component appears to become optically thin above approximately 8.4\,GHz, since the spectral index steepens again to $\alpha_{8.6}^{17} = -0.4$. This is supported by the fact that \citet{edelson87} measured a slightly steeper spectral index of $\alpha_{5}^{20} = -0.54$ between 5\,GHz and 20\,GHz for Mrk\,590 from contemporaneous observations in the year 1983.

This non-thermal radio emission at sub-pc scales could originate from a compact radio jet that is unresolved even in the VLBA images. If the emission originates from a jet, the optically thin component of the radio SED (above 8.4\,GHz) may correspond to emission from the base of the jet. The presence of a radio jet in Mrk\,590 would also be consistent with the hypothesis that low-luminosity AGNs and LINERs are accreting in the hard state \citep[][see also Section~\ref{whatishappening} of this present paper]{nagaretal01}, and as such would contain a jet as seen in hard-state Galactic black hole binaries \citep{fender01}. \citet{guptaetal15} detect high-velocity ($\sim 0.1c$) X-ray absorbers in Mrk\,590 at distances of $10^{-4}$\,pc from the black hole, indicative of ultra-fast outflows. They speculate that these outflows could be a `failed jet' (i.e., a jet that dissipates before it leaves the central few pc), arguing that the velocities are too large (relative to the observed luminosities) to be winds driven solely by radiation pressure. 

Self-absorbed non-thermal emission can also be produced by magnetized coronal winds originating from the AGN accretion disk \citep{laorbehar08}. In coronally active stars, there is a strong correlation between the radio luminosity, $L_{\rm R}$, and X-ray luminosity, $L_{\rm X}$, where ${\rm log}(L_{\rm R}/L_{\rm X}) \sim -5$ \citep{gudelbenz93}. Due to similar correlations observed by \citet{laorbehar08} in the Palomar-Green (PG) radio-quiet quasar sample and by \citet{beharetal15} in a sample of seven radio-quiet Seyfert galaxies, these authors propose that both the radio and X-ray emission in radio-quiet AGNs arise from activity in the disk-corona. Radio-loud quasars typically have higher values of ${\rm log}(L_{\rm R}/L_{\rm X}) \gtrsim -4.5$ relative to radio-quiet quasars \citep[where ${\rm log}(L_{\rm R}/L_{\rm X}) \lesssim -4.5$,][]{terashimawilson03}, likely due to contributions from the extended jet emission in addition to the coronal emission in the former. \citet{panessaetal07} find a similar dichotomy between nearby radio-quiet Seyferts and low-luminosity radio galaxies. We derive the values of $L_{\rm R}/L_{\rm X}$ for Mrk\,590, using the 2--10 keV X-ray fluxes as measured by EXOSAT in 1984 and \textit{Swift} in 2015, as well as the corresponding 5\,GHz radio flux densities with the minimum time gap relative to these two X-ray measurements (see Table~\ref{Xrayfluxes}). We find that in the 1980s and in 2015, ${\rm log}(L_{\rm R}/L_{\rm X}) \sim -5$, consistent with the radio emission in Mrk\,590 originating from coronal activity. However, we cannot rule out the presence of a weak jet, since some AGNs with ${\rm log}(L_{\rm R}/L_{\rm X}) \lesssim -5$ do contain resolved jet-like features in their radio images \citep{panessagiroletti13}. It is also entirely possible that the core radio emission in Mrk\,590 originates from both coronal winds and jets. In fact, these two forms of outflows may even be equivalent and indistinguishable in Seyfert galaxies with unresolved compact cores \citep{kharbetal15}. Some models, for example, postulate that the disk-corona itself forms the base of the jet \citep{beloborodov99,merlonifabian02}. 

Higher angular resolution VLBI observations at higher frequencies can provide stronger constraints on the size of the radio-emitting region, and determine if there are sub-pc scale jets in Mrk\,590.

\begin{table*}
\centering
\caption{X-ray fluxes and accretion properties of Mrk\,590. \label{Xrayfluxes}}
\begin{tabular}{lcccccc}
\hline
\hline
Year & Instrument & Energy range & ${F_{\rm X}}^a$ & Ref. & ${L_{\rm bol}/L_{\rm Edd}}^b$ & ${{\rm log_{10}}(L_{\rm R}/L_{\rm X)}}^c$ \\
\hline
1984 & EXOSAT & 2--10 keV & $27.0 \pm 2.7$ & 1 & 0.061 & $-$5.2 \\
2015 & \textit{Swift} & 2--10 keV & $2.60 \pm 0.55$ & 2 & 0.006 & $-$4.3 \\
\hline
\end{tabular}
\begin{flushleft}
$^a$X-ray continuum flux in units of $10^{-12}\,\rm erg\,s^{-1}\,cm^{-2}$\\
$^b$estimated using $L_{\rm bol} = 10L_{\rm X}$ \citep{vasudevanfabian07}. The uncertainties are not included as these are crude estimates.\\
$^c$The values of $L_{\rm R} = \nu L_{\nu}$ are derived using the 5\,GHz radio flux densities measured by the VLA in 1982 and in 2015. The uncertainties are not included as these are rough estimates.\\
References. -- (1)~\citet{turnerpounds89}. (2) Lawther et al. (in prep).
\end{flushleft}
\end{table*}

\subsection{Nuclear radio variability}\label{radiovar}

As described by \citet{denneyetal14} and summarised in Section~\ref{clmrk590}, Mrk\,590 exhibited an order of magnitude increase in the optical-UV continuum, broad-line (H$\beta$) fluxes from the 1970s to the early 1990s. The optical-UV (both continuum and line) fluxes then decreased by two orders of magnitude or more from the 1990s to the present. We now investigate if we observe any radio flux variability in Mrk\,590 over the same period.

\subsubsection{Is Mrk\,590 variable at radio frequencies?}\label{}

We show the archival, published and newly obtained radio flux densities of Mrk\,590 at 1.4\,GHz, 4.9\,GHz and 8.4\,GHz as a function of time in Figure~\ref{radiolightcurve}. For the 2015 VLA observations, we select the basebands for which the centre frequencies most closely match that of the earlier observations, i.e., 1.3\,GHz, 5.0\,GHz and 8.6\,GHz. Interpreting the radio flux variability of Mrk\,590 from this figure is complicated by the wide range of beam sizes (i.e. angular resolutions) achieved with different interferometers in different array configurations. In fact, some of the flux variations can be attributed to larger beams (shown as larger symbols) with higher surface brightness sensitivities picking up more emission undetected in higher angular resolution observations. There are, however, hints of real astrophysical variability in the radio lightcurves of Mrk\,590 consistent with the pattern of variability observed at optical-UV and X-ray frequencies over the past four decades:

\begin{figure}
\begin{center}
\includegraphics[width=\columnwidth]{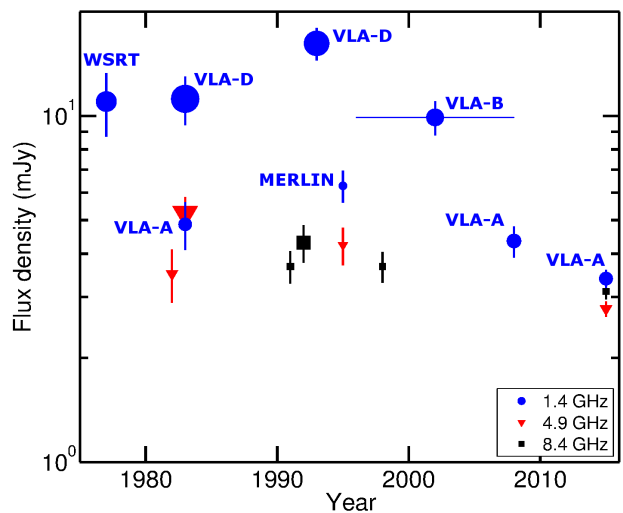}
\end{center}
\caption{Multi-frequency archival and newly-observed radio core flux densities measured from the 1970s to 2015 by the WSRT, MERLIN, and VLA. The sizes of the individual symbols scale (logarithmically) with the FWHM diameter of the clean beam of each observation. The instruments (plus array configurations for the VLA) responsible for obtaining each measurement at 1.4\,GHz are also explicitly shown to improve clarity. {\label{radiolightcurve}}}
\end{figure}

\begin{enumerate}
 
\item The 1.4\,GHz flux density increases by $28\%$, from 4.9\,mJy as measured by the VLA in A-configuration in 1983 to 6.3\,mJy as measured by MERLIN in 1995. This is in spite of MERLIN having a much smaller beam size (i.e., lower sensitivity to extended, low surface brightness components) than the VLA in A-configuration (see Table~\ref{archivaldata}). There also appears to be a similar increase in flux density between the years 1983 and 1993 as observed by the VLA at much lower angular resolutions in D-configuration (large circle symbols in Figure~\ref{radiolightcurve}). 
 
\item There is a $29\%$ decrease in flux density between 1.4\,GHz MERLIN observations in 1995 and 1.4\,GHz observations in 2008 obtained from the VLAS82 catalogue \citep{hodgeetal11}. Comparing the same MERLIN observations with that of our 2015 VLA observations at 1.3\,GHz, we measure a $46\%$ ($\sim 4 \sigma$) flux decrease. Again, this is in spite of the larger beam sizes of the 2008 and 2015 VLA A-configuration observations relative to the MERLIN beam size. 
 
\item At 4.9\,GHz, we observe a $20\%$ increase in the flux densities between 1982 and 1995, and a 34\% ($\sim 3 \sigma$) decrease in the flux densities between 1995 to 2015. Resolution effects can be ruled out, since the flux measurement in 1995 was obtained using MERLIN with a smaller beam size (even with uv-tapering applied) compared to the VLA measurements in 1982 and 2015. 
 
\item The 8.4\,GHz flux densities show a $13\%$ decrease between the years 1998 and 2015, with both measurements having comparable beam sizes. We note that this decrease is insignificant when considering our adopted conservative uncertainties, and the data are consistent with there being no variability at the $1\sigma$ level.

\item At 1.4\,GHz and 4.9\,GHz frequencies, the 2015 VLA flux densities are generally lower than the VLA flux densities obtained in the 1980s (both obtained in the A-configuration and thus have comparable beam sizes). This is consistent with the present optical-UV fluxes being lower than corresponding fluxes measured in the 1980s, when the fluxes were increasing in strength. This can also be seen in Figure~\ref{radiosed}, where the 2015 VLA flux densities are typically lower than the flux densities measured in the 1980s and 1990s.
\end{enumerate} 
 
Overall, we observe a slight increase in the radio flux densities between the 1980s and 1990s, and larger amplitude decreases in the flux densities between the 1990s to the present, consistent with the trends observed at optical-UV and X-ray frequencies. While the flux variations at 1.4\,GHz are significant, the variations at 4.9\,GHz and 8.4\,GHz are either marginal or insignificant when considering our conservative estimates of the flux uncertainties (Section~\ref{obsdata}); the significance of these variations may be higher if the systematic calibration uncertainties are overestimated, very likely the case at 4.9\,GHz and 8.4\,GHz. In any case, the correlated variations at these three radio frequencies suggests that the variability is real. 

We note that Mrk\,590 was missed in a search for variable radio sources in SDSS Stripe 82 \citep{hodgeetal13}, through comparisons of 1.4\,GHz flux densities in the FIRST and the VLAS82 catalogues. Their search criteria covers unresolved sources with flux density variations $> 5\%$, both of which are fulfilled by Mrk\,590. They, however, include in their list of variable sources only those in which the flux density increases between the earlier FIRST survey measurements and the VLAS82 measurements in 2008 to avoid contamination by variability induced by the smaller VLAS82 beam size relative to the FIRST survey. The flux density of Mrk\,590 was decreasing between the FIRST and VLAS82 epochs.

\subsubsection{Instrumental biases, interstellar scintillation or intrinsic variability?}\label{realornot}

We examine here whether the overall lower flux densities seen in our 2015 VLA observations relative to earlier observations are due to biases arising from a combination of flux averaging over larger bandwidths and non-zero source spectral indices. The 1.4\,GHz continuum flux densities may also be biased towards higher values in smaller bandwidth observations (100\,MHz for historical VLA data) relative to larger bandwidth observations (512\,MHz in the 2015 VLA observations) due to the presence of narrow HI line emission. We examined the 2015 VLA flux densities obtained using bandwidths and centre frequencies similar to that of the earlier archival datasets, and found that they were still lower in amplitude. Comparing the 2015 VLA flux densities obtained using these smaller `matched bandwidths' with those averaged over larger bandwidths presented in Table~\ref{VLAnewfluxes} reveals no significant differences (less than a few percent) in their values. As can be seen in Figure~\ref{radiosed}, the spectral shape is relatively flat between 1.4 to 8.4\,GHz, and the temporal flux variations are much larger than the expected spectral flux variations within the 1\,GHz bandwidths of the observations. Therefore, the lower flux densities as derived from our 2015 VLA observations cannot be attributed to the use of different bandwidths at different epochs and HI contamination at 1.4\,GHz in older flux measurements. 

Although we cannot definitively rule out interstellar scintillation (ISS), we argue that it is unlikely to be the cause of the observed radio continuum variations. At mid-Galactic latitudes ($b = -55^{\circ}$ for Mrk\,590) compact extragalactic radio sources are known to exhibit strong refractive ISS at frequencies $\lesssim 1$ GHz \citep{rickett86}. If the $\sim 30\%$ to $46\%$ variations observed in Mrk\,590 at 1.4\,GHz are due to strong refractive ISS, a compact component with an angular size of $\sim 50\, {\rm \mu as}$ (0.02\,pc) would be required, following the equations given by \citet{walker98}. However, the expected scintillation timescales would be of order $\sim 1$\,day for such a compact source, rather than on the $\sim 10$ year timescales seen here. Although it is possible that such a compact component of Mrk\,590 could in fact be scintillating at these timescales, and was missed due to the sparse time-sampling, it is highly unlikely considering the similarity in trends observed in the radio and optical-UV continuum lightcurves. We also do not see any single data point in Figure~\ref{radiolightcurve} exhibiting variability contrary to that expected from resolution effects and the slow increase and decrease consistent with the optical-UV continuum, which would require variability arising from stochastic processes on shorter timescales (days to months) to be invoked. At 5\,GHz to 8\,GHz, weak ISS dominates the radio variability of compact $\mu$as radio sources on timescales of hours, days, and weeks, but the typically $\lesssim 10\%$ variations \citep{lovelletal08,koayetal11} due to weak ISS are smaller than the 20\% to 30\% variations observed in Mrk\,590 at 4.9\,GHz. Follow-up monitoring of the intra and inter-day variability of Mrk\,590 at these frequencies will be needed to rule out (or confirm) the presence of ISS with these amplitudes. Such observations can also place further constraints on the size of the radio emitting region \citep[e.g.,][]{walker98}.  

The decades-long timescale of the observed radio variations is consistent with intrinsic variability of a source no larger than 3 parsecs, based on light-travel time arguments, and as corroborated by the VLBA images of the compact radio core. While the radio variability (particularly at 1.4\,GHz) shows trends that are similar to that observed at optical-UV and X-ray frequencies, also suggesting an origin intrinsic to the source, these radio flux variations are marginal in comparison to the order of magnitude flux variations observed at the higher optical-UV and X-ray frequencies. Radio flux variations at similarly low (20 to 50\%) levels have been observed on timescales of weeks to years in some nearby Seyfert galaxies \citep{mundelletal09}, including NGC\,5548 \citep{wrobel00} and NGC\,4051 \citep{jonesetal11}. In the case of NGC\,4051, the radio variability is also much lower in amplitude than the variability observed in X-rays \citep{jonesetal11}. The percentage of radio variability in Mrk\,590 could be higher if the variable component is responsible for only a fraction of the total flux measured by the VLA and the VLBA.

\subsubsection{What is happening in Mrk\,590?}\label{whatishappening}

The similarity in trends between the radio flux variations and the line and continuum variations observed at optical-UV and X-ray frequencies provides further evidence that the changing-look behaviour in Mrk\,590 is caused by variable accretion, rather than obscuration by an intervening cloud of gas and dust. It also confirms the physical connection between the radio-emitting outflow (either a jet or a coronal wind) and the accretion process. 

The increase in accretion rate between the 1980s and 1990s may have triggered a flare or the temporary acceleration of the bulk flow, producing internal shocks within the AGN jet or wind \citep[e.g.,][]{fenderetal04,jamiletal10}, that subsequently expanded and faded as the accretion rates decreased. We do not significantly detect any other `blobs' in our VLBA images in addition to the central core component, that may confirm the ejection of a new component by the nucleus. This may be due to the fact that Seyfert jets and outflows are known to be typically non-relativisitic \citep[e.g.,][]{middelbergetal04}, such that new components may not have propagated beyond a distance of 1\,pc within the past few decades. The outflow could also be aligned close to the line of sight, such that the proper motions are negligible.  

For an optically thick source, one expects to observe stronger flux variations at higher frequencies which probe the innermost regions of the emitting region. Intriguingly, the radio flux density variations in Mrk\,590 are most prominent at 1.4\,GHz, and are less significant at higher frequencies. This could be caused by the emergence and fading of a new outflowing component that is optically thin and has a steep spectrum, with a spectral turnover frequency at $\lesssim 1$\,GHz. This is consistent with the steeper spectral index of $\alpha_{1.6}^{8.4} = - 0.32$ measured in the 1990s \citep{theanetal01}, compared to the recently measured flat spectral index of $\alpha_{1.8}^{8.4} = 0.03$ in the present. Another possible explanation is the increase in synchrotron self-absorption between the 1990s and the present, causing the large decrease in flux at lower frequencies. We note also that the 8.4\,GHz flux density was measured in 1998, whereas the flux densities at 1.4\,GHz and 5\,GHz were obtained in 1995. The 8.4\,GHz flux density may therefore already have decreased further by 1998, resulting in lower variability amplitudes when compared to the 2015 measurements (more so if the 8.4\,GHz lightcurve leads the 1.4\,GHz lightcurve due to opacity effects). 

\begin{figure}
\begin{center}
\includegraphics[width=\columnwidth]{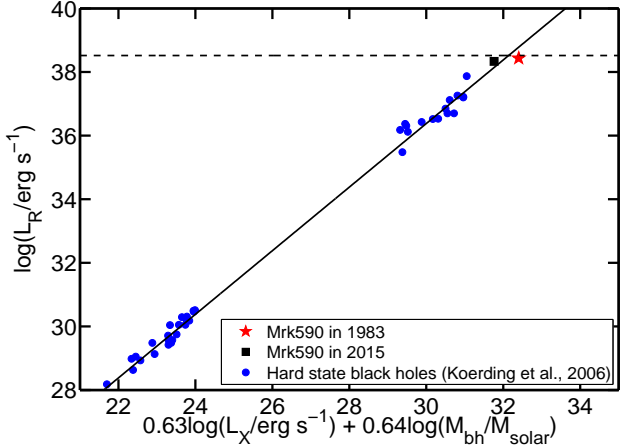}
\end{center}
\caption{The best-fit fundamental plane of black hole activity (solid line) as given by \citet{kordingetal06}, for the sample containing only Galactic black holes accreting in the low/hard state and LINERs (circles). We include Mrk\,590 when it was accreting at a higher rate in 1983 (star symbol), and in its low 2015 state (square symbol). Due to the lack of 2--10\,keV X-ray flux measurements in the 1990s, we show only the radio luminosity of Mrk\,590 observed in 1995 (dashed horizontal line). {\label{funplane}}}
\end{figure}

One interesting question to ask, is whether the variability of Mrk\,590 may be analogous to any of the different types of variability observed in Galactic black hole binaries, e.g., correlated X-ray and radio variability within the low/hard state, or a state transition between radiatively efficient (high/soft) accretion into radiatively inefficient (low/hard) accretion \citep[e.g.,][and references therein]{fendergallo14}. The best-fit fundamental plane of black hole activity as derived using a sample containing only hard state Galactic black holes and LINERs \citep{kordingetal06} is shown in Figure~\ref{funplane}, with the observed values for Mrk\,590 also added to the figure. The X-ray and radio luminosities of Mrk\,590 during the 1980s and in the present lie close to this relationship (within the 0.12 dex intrinsic scatter of the observational data), consistent with hard state accretion in both of these epochs. The correlated radio and X-ray variability over the past few decades, as described in Section~\ref{radiovar}, is also typical of black holes in the hard state, for which both the X-ray and radio fluxes are believed to be dominated by emission from a quasi-steady jet.

The high accretion rate of $L_{\rm bol}/L_{\rm Edd} \sim 0.1$ \citep{petersonetal04} and the presence of a strong power-law continuum component in the optical-UV spectra in the 1990s \citep{denneyetal14}, suggest that the AGN may have transitioned into the high/soft state temporarily at the time. This optical-UV continuum emission from the AGN \citep[the so-called `big blue-bump',][]{sandersetal89} is often associated with thermal emission from a geometrically thin accretion disk \citep{shakurasunyaev73}, observed as soft X-ray emission in Galactic black holes in the high/soft state. The nearly horizontal trajectory of Mrk\,590 in Figure~\ref{funplane} between the 1980s and the present, due to the stronger flux decrease in X-rays relative to that of the radio flux, also hints at this possibility. Unfortunately, there are no known measurements of the 2--10\,keV flux of Mrk\,590 in the 1990s when it was accreting at its peak. If the X-ray flux then was significantly higher than that in the 1980s, Mrk\,590 may have deviated significantly from the fundamental plane relationship and become radio-weak relative to the X-ray luminosity, as expected for high/soft state accretion. This is because the radio luminosity did not increase significantly (dashed horizontal line in Figure~\ref{funplane}). In such a scenario, the higher amplitude radio flux densities in the 1990s may be associated with a bright radio outburst or flare, similar to that observed in Galactic black hole binaries during transitions from low/hard accretion states to high/soft states \citep[e.g.,][]{fenderetal99,fenderetal04}. The flux measurements in the 1990s may have been obtained at the tail end of the flare, such that it has evolved from being optically thick to being optically thin \citep{vanderlaan66} as a result of expansion, explaining the more significant flux differences at 1.4\,GHz relative to the higher frequencies. If such a state transition occurred, such that the outflows are also transitioning from being jet-dominated to being wind-dominated \citep[e.g.,][]{fenderetal99}, it may explain why the observational data are consistent with the emission arising from both coronal winds or jets. However, the expected timescale of such state transitions are $\sim 10^5$ years when scaled linearly from state transitions in Galactic black holes \citep{sobolewskaetal11}, inconsistent with the timescale of the changing-look behaviour in Mrk\,590. The scaling of variability timescales between stellar mass black holes and supermassive black holes would have to be non-linear if indeed state transitions in AGNs occur on such short timescales (as also noted by \citet{schawinskietal10}), contrary to that suggested by other studies \citep[e.g.,][]{mchardyetal06}. We note that \citet{vanvelzenetal16} recently observed a possible accretion state transition on a timescale of a few months during the tidal disruption event (ASASSN-14li) of a supermassive black hole.

Should the accretion rate of the AGN in Mrk\,590 increase significantly again, our baseline measurements will be crucial for comparisons with future X-ray and radio data to examine if this AGN indeed transitions between different modes of accretion. Flux monitoring with better time sampling will provide better lightcurves for determining the relative time-lags between components emitting at different frequencies to constrain the relative sizes of the emission regions. Better frequency sampling of multi-epoch data will provide a clearer picture of the spectral changes and possibly constrain physical models describing the flux variations. Follow-up imaging with VLBI (5 to 10 years from now) may reveal possible changes in the parsec-scale morphology, e.g., resulting from newly ejected synchrotron emitting components travelling beyond the central 1\,pc.
                	               
\section{Summary}\label{summary}

We have conducted the first investigation into the radio properties of a changing-look AGN by examining the parsec-scale morphology, spectral shape, and variability of the radio continuum emission in Mrk\,590, a nearby Seyfert galaxy that has undergone a significant decline in accretion rate during the past decade. Our main results can be summarised as follows: 
\begin{enumerate}
\item We detect a parsec-scale radio core component with a flux density of $\approx 3\,\rm mJy$ at both 1.6\,GHz and 8.4\,GHz. The emission detected on hundreds of pc to kpc scales in VLA images from 1.8\,GHz to 17\,GHz is dominated by emission from this component. 

\item The flat spectrum ($\alpha_{1.6}^{8.4} = 0.03$) and high brightness temperature ($T_{\rm b} \sim 10^8 \rm \,K$ ) of this compact radio emitting region indicates non-thermal, synchrotron emission from multiple, optically thick source components. The radio to X-ray luminosity ratio of ${\rm log}(L_{\rm R}/L_{\rm X}) \sim -5$ in Mrk\,590, similar to that observed in coronally active stars, suggests that magnetized coronal winds may be responsible for the radio emission, although a compact radio emitting jet is also consistent with the data. The high brightness temperatures also rule out supernova remnants and HII regions, as well as thermal emission from ADAFs or the inner part of the AGN torus, as the dominant source of radio emission in Mrk\,590.

\item We discover a $28\%$ flux increase between the years 1983 and 1995 and a $46\%$ flux decrease between 1995 and 2015 at 1.4\,GHz. These variations are also seen (marginally, at 13\% to 30\% levels) at 5\,GHz and 8.4\,GHz. They are consistent with trends observed at optical-UV and X-ray wavelengths. Such correlated variability provides further evidence that the changing-look behaviour in Mrk\,590 is caused by variations in the accretion properties as opposed to obscuration of the accretion disk and BLR by dusty clouds. The radio flux variations may be caused by a flare as a result of the increased accretion rate, leading to the emergence of shocks within the outflow (i.e. jet or wind), that eventually expanded and faded in luminosity as the accretion rate declined. 

\item The radio and X-ray luminosities of Mrk\,590 in the 1980s and in 2015 lie within the 0.12\,dex observational scatter of the fundamental plane of black hole activity \citep{kordingetal06} for objects accreting in the low/hard state. The AGN may have transitioned briefly into the high/soft state in the 1990s when it was accreting at its peak ($\sim$10\% of the Eddington limit), as evidenced by the presence of an optical-UV continuum component \citep{denneyetal14} associated with a thin accretion disk.  

\end{enumerate}

This study highlights the importance of including radio data in multiwavelength observations of changing-look AGNs, and more generally for studies of variable accretion in AGNs. However, the sparse time sampling of available archival data for Mrk\,590 precludes more detailed studies of the coupling between the various physical components (BLR, accretion disk, corona, and outflows) emitting at different wavelengths. Therefore, high cadence multi-wavelength monitoring of AGN variability, such as radio and X-ray monitoring in support of decades-long reverberation mapping programmes \citep[e.g.,][]{kingetal15,shenetal15}, will provide better characterizations of the time delays between the radio, optical-UV and X-ray lightcurves in a large sample of sources, including that of any changing-look AGNs that may be uncovered.

\section*{Acknowledgements}

We express our sincere gratitude to Sadie Jones for providing us with the data from the paper by \citet{kordingetal06} which were used in Figure~\ref{funplane} of this paper. We are also grateful to Daniel Lawther for providing the (as yet unpublished) \textit{Swift} X-ray fluxes of Mrk\,590 measured in the year 2015. We thank Paul Ho and Sandra Raimundo for the valuable discussions, as well as the anonymous reviewer for the helpful comments. We also thank Anita Richards for her help in providing the MERLIN visibility data. We are grateful to the staff at the National Radio Astronomy Observatory (NRAO), particularly Heidi Medlin, for advice and assistance with the scheduling of the VLA and VLBA observations. This research is partly supported by a research grant (VKR023371) from Villumfonden. MV and JYK gratefully acknowledge support from the Danish Council for Independent Research via grant no. DFF 4002-00275. The Dark Cosmology Centre is funded by the Danish National Research Foundation. BMP is grateful to the US National Science Foundation (NSF) for support through grant AST--1008882 to The Ohio State University. The NRAO is a facility of the NSF operated under cooperative agreement by Associated Universities, Inc. MERLIN is a National Facility operated by the University of Manchester at Jodrell Bank Observatory on behalf of STFC. We also acknowledge the use of data from the NASA Extragalactic Database (NED), the VLA Data Archive, and the MERLIN Data Archive.




\begin{thebibliography}{99}


\bibitem[Antonucci(1993)]{antonucci93} Antonucci R., 1993, \araa, 31, 473

\bibitem[Aretxaga et al.(1999)]{aretxagaetal99} Aretxaga I., Joguet B., Kunth D., Melnick J., Terlevich R.~J., 1999, \apjl, 519, L123

\bibitem[Baldwin et al.(1981)]{baldwinetal81} Baldwin J.~A., Phillips M.~M., Terlevich R., 1981, \pasp, 93, 5 

\bibitem[Becker et al.(1995)]{beckeretal95} Becker R.~H., White R.~L., Helfand, D.~J., 1995, \apj, 450, 559

\bibitem[Behar et al.(2015)]{beharetal15} Behar E., Baldi R.~D., Laor A., Horesh. A., Stevens J., Tzioumis T., 2015, \mnras, 451, 517

\bibitem[Beloborodov(1999)]{beloborodov99} Beloborodov A.~M., 1999, \apjl, 510, L123

\bibitem[Bianchi et al.(2005)]{bianchietal05} Bianchi S., Guainazzi M., Matt G. et al., Chiaberge M., Iwasawa K., Fiore F., Maiolino R., 2005, \aap, 442, 185

\bibitem[Blundell \& Kuncic(2007)]{blundellkuncic07} Blundell K.~M., Kuncic Z., 2007, \apjl, 668, L103

\bibitem[Cohen et al.(1986)]{cohenetal86} Cohen R.~D., Puetter R.~C., Rudy R.~J., Ake T.~B., Foltz C.~B., 1986, \apj, 311, 135 

\bibitem[Condon(1992)]{condon92} Condon J.~J., 1992, \araa, 30, 575

\bibitem[Condon et al.(1998)]{condonetal98} Condon J.~J., Cotton W.~D., Greisen E.~W., Yin Q. F., Perley R. A., Taylor G. B., Broderick J. J., 1998, \aj, 115, 1693

\bibitem[Cotton(2008)]{cotton08} Cotton W.~D.\ 2008, \pasp, 120, 439 

\bibitem[Denney et al.(2014)]{denneyetal14} Denney K. D. et al., 2014, \apj, 796, 134

\bibitem[Edelson(1987)]{edelson87} Edelson R. A., 1987, \apj, 313, 651

\bibitem[Elitzur(2012)]{elitzur12} Elitzur M., 2012, \apjl, 747, L33

\bibitem[Elitzur \& Shlosman(2006)]{elitzurshlosman06} Elitzur M., Shlosman I., 2006, \apjl, 648, L101

\bibitem[Elitzur et al.(2014)]{elitzuretal14} Elitzur M., Ho L.~C., Trump J.~R., 2014, \mnras, 438, 3340

\bibitem[Falcke et al.(1996)]{falckeetal96} Falcke H., Patnaik A.~R., Sherwood W., 1996, \apjl, 473, L13

\bibitem[Falcke et al.(2004)]{falckeetal04} Falcke,H., K{\"o}rding E., Markoff S., 2004, \aap, 414, 895

\bibitem[Fender(2001)]{fender01} Fender R.~P., 2001, \mnras, 322, 31

\bibitem[Fender et al.(1999)]{fenderetal99} Fender R. et al., 1999, \apjl, 519, L165

\bibitem[Fender et al.(2004)]{fenderetal04} Fender R.~P., Belloni T.~M., Gallo E.\ 2004, \mnras, 355, 1105

\bibitem[Fender \& Gallo(2014)]{fendergallo14} Fender R., Gallo E., 2014, \ssr, 183, 323

\bibitem[Gabor \& Bournaud(2013)]{gaborbournaud13} Gabor J.~M., Bournaud F., 2013, \mnras, 434, 606

\bibitem[Gallimore et al.(1997)]{gallimoreetal97} Gallimore J.~F., Baum S.~A., O'Dea, C.~P., 1997, \nat, 388, 852

\bibitem[Greisen(2003)]{greisen03} Greisen E.~W., 2003, Information Handling in Astronomy - Historical Vistas, 285, 109

\bibitem[G\"{u}del \& Benz(1993)]{gudelbenz93} G\"{u}del M., Benz A.~O., 1993, \apjl, 405, L63

\bibitem[Gupta et al.(2015)]{guptaetal15} Gupta A., Mathur S., Krongold Y., 2015, \apj, 798, 4 

\bibitem[Hodge et al.(2011)]{hodgeetal11} Hodge J. A., Becker R. H., White R. L., Richards G. T. , Zeimann, G. R., 2011, \aj, 142, 3

\bibitem[Hodge et al.(2013)]{hodgeetal13} Hodge J.~A., Becker R.~H., White R.~L., Richards G.~T., 2013, \apj, 769, 125

\bibitem[Jamil et al.(2010)]{jamiletal10} Jamil O., Fender R.~P., Kaiser C.~R., 2010, \mnras, 401, 394

\bibitem[Jones et al.(2011)]{jonesetal11} Jones S., McHardy I., Moss D., Seymour N., Breedt E., Uttley P., K{\"o}rding E., Tudose V., 2011, \mnras, 412, 2641 

\bibitem[Keel et al.(2015)]{keeletal15} Keel W.~C. et al., 2015, \aj, 149, 155 

\bibitem[Kewley et al.(2006)]{kewleyetal06} Kewley L.~J., Groves B., Kauffmann G., Heckman T., 2006, \mnras, 372, 961

\bibitem[Kharb et al.(2015)]{kharbetal15} Kharb P., Das M., Paragi Z., Subramanian S., \& Chitta L.~P., 2015, \apj, 799, 161

\bibitem[King et al.(2015)]{kingetal15} King A.~L. et al., 2015, \mnras, 453, 1701

\bibitem[Kinney et al.(2000)]{kinneyetal00} Kinney A.~L., Schmitt H.~R., Clarke C.~J., Pringle J. E., Ulvestad J. S., Antonucci R. R. J., 2000, \apj, 537, 152

\bibitem[Koay et al.(2011)]{koayetal11} Koay J.~Y. et al., 2011, \aj, 142, 108

\bibitem[Koay et al.(2016)]{koayetal16} Koay J.~Y., Vestergaard M., Casasola V., Lawther D., Peterson B.~M., 2016, \mnras, 455, 2745

\bibitem[K{\"o}rding et al.(2006)]{kordingetal06} K{\"o}rding E., Falcke H., Corbel S.\ 2006, \aap, 456, 439

\bibitem[Kovalev et al.(2008)]{kovalevetal08} Kovalev Y.~Y., Lobanov A.~P., Pushkarev A.~B., Zensus J.~A., 2008, \aap, 483, 759

\bibitem[Kukula et al.(1995)]{kukulaetal95} Kukula M.~J., Pedlar A., Baum S.~A., O'Dea C.~P., 1995, \mnras, 276, 1262  

\bibitem[LaMassa et al.(2015)]{lamassaetal15} LaMassa S.~M. et al., 2015, \apj, 800, 144

\bibitem[Laor \& Behar(2008)]{laorbehar08} Laor A., Behar E.\ 2008, \mnras, 390, 847

\bibitem[Lin \& Shields(1986)]{linshields86} Lin D.~N.~C., Shields G.~A., 1986, \apj, 305, 28

\bibitem[Lovell et al.(2008)]{lovelletal08} Lovell J.~E.~J. et al., 2008, \apj, 689, 108

\bibitem[MacLeod et al.(2015)]{macleodetal15} MacLeod C.~L. et al., 2015, arXiv:1509.08393

\bibitem[McHardy et al.(2006)]{mchardyetal06} McHardy I.~M., Koerding E., Knigge C., Uttley P., Fender R.~P.\ 2006, \nat, 444, 730

\bibitem[McMullin et al.(2007)]{mcmullinetal07} McMullin J.~P., Waters B., Schiebel D., Young W., Golap K., 2007, Astronomical Data Analysis Software and Systems XVI, 376, 127

\bibitem[Mahadevan(1997)]{mahadevan97} Mahadevan R., 1997, \apj, 477, 585  

\bibitem[Marchese et al.(2012)]{marcheseetal12} Marchese E., Braito V., Della Ceca R., Caccianiga A., Severgnini, P., 2012, \mnras, 421, 1803

\bibitem[Merloni \& Fabian(2002)]{merlonifabian02} Merloni A., Fabian A.~C.\ 2002, \mnras, 332, 165

\bibitem[Merloni et al.(2003)]{merlonietal03} Merloni A., Heinz S., di Matteo T., 2003, \mnras, 345, 1057

\bibitem[Merloni et al.(2015)]{merlonietal15} Merloni A. et al. \ 2015, \mnras, 452, 69

\bibitem[Middelberg et al.(2004)]{middelbergetal04} Middelberg E., et al., 2004, \aap, 417, 925

\bibitem[Miller et al.(1993)]{milleretal93} Miller P., Rawlings S., Saunders R., 1993, \mnras, 263, 425

\bibitem[Mundell et al.(2000)]{mundelletal00} Mundell C.~G., Wilson A.~S., Ulvestad J.~S., Roy A.~L., 2000, \apj, 529, 816

\bibitem[Mundell et al.(2009)]{mundelletal09} Mundell C.~G., Ferruit P., Nagar N., Wilson A.~S., 2009, \apj, 703, 802

\bibitem[Nagar et al.(2001)]{nagaretal01} Nagar N.~M., Wilson A.~S., Falcke H., 2001, \apjl, 559, L87

\bibitem[Narayan et al.(1998)]{narayanetal98} Narayan R., Mahadevan R., Quataert E., 1998, Theory of Black Hole Accretion Disks, ed. Abramowicz M. A., Bjornsson G., Pringle J. E., Cambridge University Press, 148

\bibitem[Nenkova et al.(2008)]{nenkovaetal08} Nenkova M., Sirocky M.~M., Nikutta R., Ivezi{\'c}, {\v Z}., Elitzur M., 2008, \apj, 685, 160 

\bibitem[Novak et al.(2011)]{novaketal11} Novak G. S., Ostriker J. P., Ciotti L., 2011, \apj, 737, 26

\bibitem[Panessa et al.(2007)]{panessaetal07} Panessa F., Barcons X., Bassani L., Cappi M., Carrera F. J., Ho L. C., Pellegrini S., 2007, \aap, 467, 519

\bibitem[Panessa \& Giroletti(2013)]{panessagiroletti13} Panessa F., Giroletti M., 2013, \mnras, 432, 1138

\bibitem[Penston \& P\'{e}rez(1984)]{penstonperez84} Penston M. V., P\'{e}rez E., 1984, \mnras, 211, 33P

\bibitem[Perley \& Butler(2013)]{perleybutler13} Perley R.~A., Butler B.~J., 2013, \apjs, 204, 19

\bibitem[Peterson(1997)]{peterson97} Peterson B.~M., 1997, An Introduction to Active Galactic Nuclei, (Cambridge, UK: Cambridge University Press)

\bibitem[Peterson et al.(2004)]{petersonetal04} Peterson B. M. et al., 2004, \apj, 613, 682

\bibitem[Remillard \& McClintock(2006)]{remillardmcclintock06} Remillard R.~A., McClintock J.~E., 2006, \araa, 44, 49

\bibitem[Rickett(1986)]{rickett86} Rickett B.~J., 1986, \apj, 307, 564 

\bibitem[Risaliti et al.(2005)]{risalitietal05} Risaliti G., Elvis M., Fabbiano G., Baldi A., Zezas A., 2005, \apjl, 623, L93

\bibitem[Risaliti et al.(2009)]{risalitietal09} Risaliti G. et al, 2009, \apj, 696, 160

\bibitem[Ruan et al.(2015)]{ruanetal15} Ruan J.~J. et al., 2015, arXiv:1509.03634

\bibitem[Runnoe et al.(2016)]{runnoeetal16} Runnoe J.~C. et al., 2016, \mnras, 455, 1691

\bibitem[Sanders et al.(1989)]{sandersetal89} Sanders D.~B., Phinney E.~S., Neugebauer G., Soifer B.~T., Matthews K., 1989, \apj, 347, 29

\bibitem[Schawinski et al.(2010)]{schawinskietal10} Schawinski K. et al., 2010, \apjl, 724, L30 

\bibitem[Schmitt et al.(2001)]{schmittetal01} Schmitt H. R., Ulvestad J. S., Antonucci R. R. J, Kinney, A. L., 2001, \apjs, 132, 199

\bibitem[Shakura \& Sunyaev(1973)]{shakurasunyaev73} Shakura N.~I., Sunyaev R.~A., 1973, \aap, 24, 337

\bibitem[Shappee et al.(2014)]{shappeeetal14} Shappee B.~J. et al., 2014, \apj, 788, 48
%
\bibitem[Shen et al.(2015)]{shenetal15} Shen Y. et al.\ 2015, \apjs, 216, 4 

\bibitem[Sobolewska et al.(2011)]{sobolewskaetal11} Sobolewska M.~A., Siemiginowska A., Gierli{\'n}ski M., 2011, \mnras, 413, 2259

\bibitem[Storchi-Bergmann et al.(1993)]{storchi-bergmannetal93} Storchi-Bergmann T., Baldwin J.~A., Wilson A.~S., 1993, \apjl, 410, L11

\bibitem[Terashima \& Wilson(2003)]{terashimawilson03} Terashima Y., Wilson A.~S.\ 2003, \apj, 583, 145

\bibitem[Thean et al.(2001)]{theanetal01} Thean A. H. C, Gillibrand T. I., Pedlar A., Kukula M. J., 2001, \mnras, 327, 369
 
\bibitem[Tohline \& Osterbrock(1976)]{tohlineosterbrock76} Tohline J.~E., Osterbrock D.~E., 1976, \apjl, 210, L117

\bibitem[Turner \& Pounds(1989)]{turnerpounds89} Turner T.~J., Pounds K.~A., 1989, \mnras, 240, 833

\bibitem[Ulvestad \& Wilson(1984)]{ulvestadwilson84} Ulvestad J.~S., Wilson A.~S.\ 1984, \apj, 285, 439

\bibitem[Ulvestad et al.(2005)]{ulvestadetal05} Ulvestad J.~S., Antonucci R.~R.~J., Barvainis R., 2005, \apj, 621, 123

\bibitem[Urry \& Padovani(1995)]{urrypadovani95} Urry C. M., Padovani P., 1995, \pasp, 107, 803

\bibitem[van der Laan(1966)]{vanderlaan66} van der Laan, H., 1966, \nat, 211, 1131 

\bibitem[van Velzen et al.(2016)]{vanvelzenetal16} van Velzen S., et al., 2016, Science, 351, 62

\bibitem[Vasudevan \& Fabian(2007)]{vasudevanfabian07} Vasudevan R.~V., Fabian A.~C., 2007, \mnras, 381, 1235

\bibitem[Walker(1998)]{walker98} Walker M.~A., 1998, \mnras, 294, 307

\bibitem[Walker(2014)]{walker14} Walker C., 2014, VLBA Scientific Memo \#37  

\bibitem[Wilson \& Meurs(1982)]{wilsonmeurs82} Wilson A.~S., Meurs E.~J.~A.,\ 1982, \aaps, 50, 217 

\bibitem[Wrobel(2000)]{wrobel00} Wrobel J.~M., 2000, \apj, 531, 716

\bibitem[Yi \& Boughn(1998)]{yiboughn98} Yi I., Boughn S.~P.\ 1998, \apj, 499, 198 

\bibitem[Zakamska \& Greene(2014)]{zakamskagreene14} Zakamska N.~L., Greene J.~E.\ 2014, \mnras, 442, 784



\end{thebibliography}





\bsp	
\label{lastpage}
\end{document}